\documentclass[preprint,showpacs,preprintnumbers,amsmath,amssymb]{revtex4}

\usepackage{amssymb}
\usepackage{amsmath}
\usepackage{float}
\usepackage{graphicx}
\usepackage{subfigure}
\usepackage{color}

\begin{document}

\title{Analyses of multi-pion Hanbury-Brown-Twiss correlations for the pion-emitting
sources with Bose-Einstein condensation}

\author{Ghulam Bary, Peng Ru, Wei-Ning Zhang\footnote{wnzhang@dlut.edu.cn}}
\affiliation{School of Physics, Dalian University of Technology, Dalian, Liaoning 116024,
China}

\begin{abstract}
We calculate the three- and four-particle correlations of identical pions in
an evolving pion gas (EPG) model with Bose-Einstein condensation.  The multi-pion
correlation functions in the EPG model are analyzed in different momentum intervals
and compared with the experimental data for Pb-Pb collisions at $\sqrt{s_{NN}}=2.76$
TeV.  It is found that the multi-pion correlation functions and cumulant correlation
functions are sensitive to the condensation fraction of the EPG sources in the low
average transverse-momentum intervals of the three and four pions.  The model results
of the multi-pion correlations are consistent with the experimental data in a
considerable degree, which gives a source condensation fraction between 16~--~47\%.
\\[1ex]
\end{abstract}

\maketitle

\section{Introduction}
Two-pion Hanbury Brown--Twiss (HBT) interferometry, also known as two-pion femtoscopy, has
been widely applied in high-energy heavy-ion collisions to study the space-time structure
of particle-emitting sources by measuring the intensity correlations of two identical pions
\cite{Gyu79,Wongbook,Wie99,Wei00,Csorgo02,Lisa05}.
Because the intensity correlations occur for chaotic particle emission and disappear for
coherent particle emission, HBT interferometry can also be used to study the source
coherence \cite{Gyu79,Wongbook,Wie99,Wei00,Csorgo02,Lisa05}.
The intercept of the two-pion correlation function at zero relative momentum is related
to the source coherence degree, although many other effects may affect the measurement
value of the intercept \cite{Gyu79,Wongbook,Wie99,Wei00,Csorgo02,Lisa05}.
As extension of two-pion interferometry, multi-pion correlation analyses are developed
and carried out in high-energy heavy-ion collisions \cite{{Liu86,Zajc87,BiyaSuzuAnd,
AndPluWei9193,Zhang9390,WNZhang,Pratt93,QHZhang95,CsorgoZimanyi97,HeiVis97,HeiZhaSug,Wiedemann98,
Csorgo98,NA44,NakamuraSeki,WA98,STAR-PRL03,MorMurNak,ALICE-PRC14,ALICE-PLB14,Gangadharan15}}.
Recently, the ALICE collaboration analyzes the three- and four-pion correlations in
$pp$, $p$-Pb, and Pb-Pb collisions at the Large Hadron Collider (LHC) \cite{ALICE-PRC16}.
A significant suppression of three- and four-pion correlations observed in Pb-Pb
collisions may arise from a considerable coherence degree of the particle-emitting sources,
which is consistent with the previous measurements of three-pion correlations in the
collisions \cite{ALICE-PRC14}.  It is of interest to explain the experimental observations
of multi-pion correlations.

In Ref. \cite{WongZhang07}, C. Y. Wong and W. N. Zhang studied the pion Bose-Einstein
condensation and the chaoticity parameter $\lambda$ in two-pion HBT interferometry for
a static boson gas source within a mean-field with harmonic oscillator potential in
high-energy heavy-ion collisions.
The model of the non-relativistic boson gas within harmonic oscillator potential
can be solved analytically \cite{WongZhang07} and be used in atomic HBT correlation
analyses \cite{NarGla99,Viana06}.
In Ref. \cite{LiuRuZhangWong14}, the chaoticity parameter $\lambda$ was investigated
in an evolving pion gas (EPG) model with Bose-Einstein condensation.  The pion gas
in this model was considered within a harmonic oscillator mean-field and expanding
in relativistic hydrodynamics \cite{LiuRuZhangWong14}.
The investigations \cite{LiuRuZhangWong14} indicate that the pion sources produced in
the Pb+Pb collisions at $\sqrt{s_{NN}}=2.76$ TeV at the LHC is partially coherent,
perhaps due to a degree of Bose-Einstein condensation.
The finite condensation decreases the chaoticity parameter $\lambda$ in the two-pion
interferometry measurements in low momentum interval of pion pair, and influences very
slightly the $\lambda$ value for the pion pair with high momenta \cite{LiuRuZhangWong14}.
In this work, we shall investigate three- and four-pion HBT correlations
in the EPG model \cite{LiuRuZhangWong14}.  We shall examine the relationship between
the condensation and the strength of the multi-pion correlations in different momentum
intervals.  We shall compare the model results of multi-pion correlation functions
with the experimental data for Pb-Pb collisions at $\sqrt{s_{NN}}=2.76$~TeV at
the LHC \cite{ALICE-PRC16}.  It is speculated that the coherent fraction of the
particle-emitting sources is between 16~--~47\%, consistent with the analysis result
for the four-pion correlations measured in the collisions \cite{ALICE-PRC16}.

The rest of this paper is organized as follows.  In Sec. II, we review the EPG model
and present the calculations of the three- and four-pion correlation functions in the
EPG model with Bose-Einstein condensation.  In Sec III, we examine the multi-pion
correlation functions in different momentum intervals.  In Sec. IV, we compare the model
results of the multi-pion correlation functions with experimental data.  Finally, we
give the summary and conclusion in Sec V.

\section{Model and multi-pion correlation formulas}
\subsection{EPG model}
As in Ref. \cite{LiuRuZhangWong14}, we consider a pion-emitting source as a relativistic
boson gas of identical pions within the time-dependent harmonic oscillator potential that
arises approximately from the mean field of the hadronic medium in high-energy heavy-ion
collisions  \cite{WongZhang07,LiuRuZhangWong14,WongZhangLiuRu15},
\begin{equation}
\label{Vrt}
V(\textbf{\emph{r}},t) = \frac{1}{2}\,m\,\omega^2(t)\, r^2 =
\frac{1}{2}\,\hbar\,\omega(t)\, \frac{r^{2}}{a^{2}(t)},
\end{equation}
where $\hbar \omega(t)$ measures the potential strength and $a(t)= \sqrt{
\hbar/m\omega(t)}$ is the characteristic length of harmonic oscillator.

Assuming that the relaxation time of the system is smaller than the source
evolution time, the expansion of the pion gas may approximately deal with
a quasi-static adiabatic process \cite{LiuRuZhangWong14}.  In this case,
the temperature $T$ and volume $V$ have the relationship $TV^{\gamma-1}=$
constant, where $\gamma$ is the ratio of the specific heats at constant
pressure and volume. For example, $\gamma=\frac{5}{3}$ for non-relativistic
monatomic gas.  We assume the characteristic length $a$ is proportional to
a parameterized source radius as in Ref. \cite{LiuRuZhangWong14}, $a=C_1 R=C_1
(R_0+\alpha t)$, where the proportional parameter $C_1$ can be determined
by the source root-mean-squared radius, $R_0$ is initial radius of the
source and $\alpha$ is a parameter related to the source average expansion
velocity.  With a hydrodynamical calculation for $R_0=6$ fm, $T_0=170$ MeV,
the model parameters $\gamma$ and $\alpha$ are fixed to be 1.627 and 0.62
\cite{LiuRuZhangWong14}, respectively.  And, the parameter $C_1$ in the model
calculations in this paper is taken to be 0.35 and 0.40 as in Ref.
\cite{LiuRuZhangWong14}.

For the identical boson gas with a fixed number of particles, $N$, and at
a given temperature $T=1/\beta$, one has
\begin{equation}
\label{N0T}
N=N_0+N_T,
\end{equation}
where, $N_0$ is the number of particles in $n=0$ state,
\begin{equation}
\label{N0}
N_0=\frac{\mathcal Z}{1-\mathcal Z},
\end{equation}
and $N_T$ is the number of the particles in $n>0$ states,
\begin{equation}
\label{NTn}
N_T=\sum_{n>0}^{\infty}\frac{g_n \mathcal Z\,e^{-\beta\tilde E_n}}
{1-\mathcal Z\,e^{-\beta\tilde E_n}},
\end{equation}
where $g_n$ is the degeneracy of the $n$-th energy level, $\mathcal Z$
is the fugacity parameter which includes the factor for the lowest energy
$\varepsilon_0$, and $\tilde E_n$ is the relative energy levels to
$\varepsilon_0$ \cite{NarGla99,WongZhang07}.  Because $N_0 \ge 0$, the values
of $\mathcal Z$ are between zero and one.  From Eqs. (\ref{N0T})~---~(\ref{NTn})
and with the energy levels of harmonic oscillator, we can calculate $\mathcal Z$
numerically for fixed $N$ \cite{WongZhang07,LiuRuZhangWong14}, and then obtain
the condensation fraction,
\begin{equation}
f_0=\frac{N_0}{N}=\frac{\mathcal Z}{(1-\mathcal Z)N}\,.
\label{def-f0}
\end{equation}

\begin{figure}[htb]
\includegraphics[width=0.80\columnwidth]{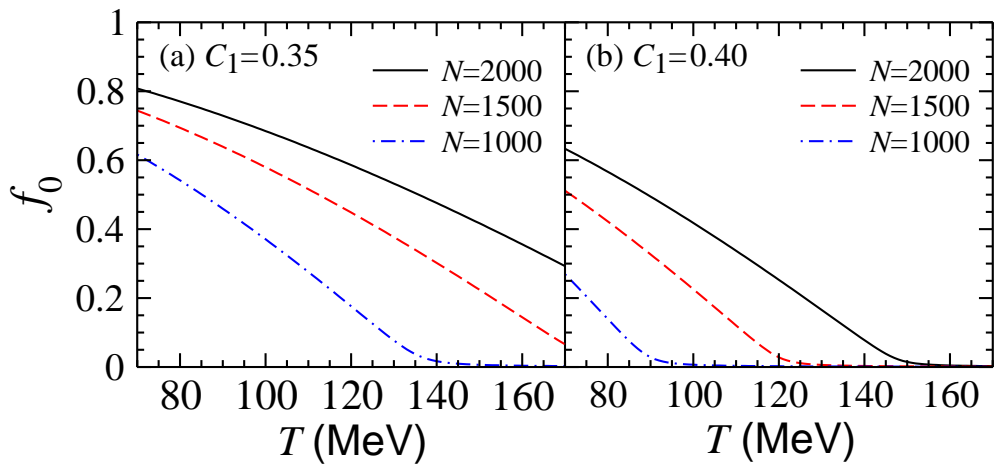}
\caption{(Color online) Condensation fraction as a function of temperature.}
\label{zf0}
\end{figure}

In Fig. \ref{zf0}, we show the condensation fractions as a function of temperature
for the sources with $N=$ 1000, 1500, and 2000.  Here, the left and right panels are
for the parameter $C_1=$ 0.35 and 0.40, respectively.
One can see that the condensation fraction $f_0$ increases with the particle number
$N$ and decreases with increasing temperature.  For fixed $N$ and $T$, the condensation
fraction for $C_1=0.35$ is higher than that for $C_1=0.40$ because the condensation is
significant for the system with a small characteristic length $a$ \cite{WongZhang07}.

As we know the density matrix of a generic quantum ensemble can be written as
\begin{equation}
{\hat \rho}=\sum_{N=0}^{\infty}{\cal P}_{\!\!_N}{\hat \rho}_{\!_N},
\end{equation}
where the set $\{{\cal P}_{\!\!_N}\}_{N=0}^{\infty}$ is normalized multiplicity
distribution, ${\hat \rho}_{_N}$ denotes the density matrix of the ensemble in which
the systems with a fixed particle number $N$, and then an observable is given by
\begin{equation}
\langle\langle\,{\hat A}\,\rangle\rangle ={\rm Tr}({\hat A}{\hat \rho})
=\sum_{N=0}^{\infty}{\cal P}_{\!\!_N} \langle\langle\,{\hat A}\,\rangle\rangle_N
=\sum_{N=0}^{\infty}{\cal P}_{\!\!_N} {\rm Tr}({\hat A}{\hat \rho}_{\!_N}),
\end{equation}
where $\langle\langle\cdots\rangle\rangle$ denotes the double average over the
quantum states of system and ensemble systems.  Quantities $\langle\langle\,{\hat A}
\,\rangle\rangle$ and $\langle\langle\,{\hat A}\,\rangle\rangle_{\!_N}$ may also be
referred to as the ``inclusive" and ``exclusive" quantities with respect to the
multiplicity of event.

In Refs. \cite{CsorgoZimanyi97}, T. Cs\"org\H{o} and J. Zim\'{a}nyi solve analytically
the multiplicity distribution, single-particle momentum spectra, and two-particle HBT
correlations using the particle-wave-packet technique, for the static identical pion
system with all order Bose-Einstein symmetrizations.  Because of the symmetrization,
the emission of pion encourages the emission of more identical pions when the particle
density is sufficiently high, which is referred to as a ``pion laser" first introduced
by S. Pratt \cite{Pratt93}.

Compared to the pion-laser model (PLM) \cite{Pratt93,CsorgoZimanyi97}, the EPG model
describes an evolving pion-emitting source.  It deals with the canonical ensemble in
which the systems of pion gas have a fixed particle number $N$ and assumed to have
certain temperature and volume at each hydrodynamically evolving state
\cite{LiuRuZhangWong14}.  Obviously, the EPG model is an approximate description for
the sources produced in high-energy heavy-ion collisions after chemical freeze-out,
and cannot be used to investigate the multiplicity distribution in the collisions.

In the EPG model, the one- and two-particle density matrices in momentum space
are \cite{LiuRuZhangWong14}
\begin{eqnarray}
\label{Gp1_1}
G^{(1)}(\textbf{\emph{p}}_1,\textbf{\emph{p}}_2)&&=\sum_n u_n^*(\textbf{\emph{p}}_1)
u_n(\textbf{\emph{p}}_2) \langle {\hat a}^{\dag}_n{\hat a}_n\rangle \nonumber\\
&&\hspace*{-18mm}=\sum_n u_n^*(\textbf{\emph{p}}_1) u_n(\textbf{\emph{p}}_2)
\frac{g_n \mathcal Z\,e^{-\beta\tilde E_n}}{1-\mathcal Z\,e^{-\beta\tilde E_n}},
\end{eqnarray}
\begin{equation}
\label{Gp1p2_1}
G^{(2)}(\textbf{\emph{p}}_1,\textbf{\emph{p}}_2;\textbf{\emph{p}}_1,
\textbf{\emph{p}}_2)=\sum_{klmn} u_k^*(\textbf{\emph{p}}_1)
u_l^*(\textbf{\emph{p}}_2) u_m(\textbf{\emph{p}}_2)
u_n(\textbf{\emph{p}}_1) \langle {\hat a}^{\dag}_k{\hat a}^{\dag}_l
{\hat a}_m{\hat a}_n\rangle,
\end{equation}
where $u_n(\textbf{\emph{p}})$ is the wave function of single-particle for the
$n$-th state, ${\hat a_n}$ (${\hat a}_n^{\dag}$) is the annihilation (creation)
operator of particle, and $\langle \cdots \rangle$ denotes the ensemble average.
The invariant single-pion momentum distribution is
\begin{equation}
E\frac{dN}{d\textbf{\emph{p}}}=\sqrt{\textbf{\emph{p}}^2+m_{\pi}^2}\,
G^{(1)}(\textbf{\emph{p}},\textbf{\emph{p}}),
\end{equation}
and the two-pion correlation function is defined as
\begin{equation}
\label{Cp1p2_1}
C_2(\textbf{\emph{p}}_1,\textbf{\emph{p}}_2)=\frac{G^{(2)}(\textbf{\emph{p}}_1,
\textbf{\emph{p}}_2;\textbf{\emph{p}}_1,\textbf{\emph{p}}_2)}
{G^{(1)}(\textbf{\emph{p}}_1,\textbf{\emph{p}}_1)\,G^{(1)}(\textbf{\emph{p}}_2,
\textbf{\emph{p}}_2)}.
\end{equation}
In the limit of a large number of particles, $N(N-1)\sim N^2(\,\gg N_T,N_0)$,
the numerator in Eq. (\ref{Cp1p2_1}) can be written as
\cite{WongZhang07,WongZhangLiuRu15}
\begin{eqnarray}
\label{Gp1p2_2}
&&G^{(2)}(\textbf{\emph{p}}_1,\textbf{\emph{p}}_2;\textbf{\emph{p}}_1,
\textbf{\emph{p}}_2)=G^{(1)}(\textbf{\emph{p}}_1,\textbf{\emph{p}}_1)
G^{(1)}(\textbf{\emph{p}}_2,\textbf{\emph{p}}_2)\nonumber\\
&&\hspace*{20mm}+G^{(1)}(\textbf{\emph{p}}_1,\textbf{\emph{p}}_2)G^{(1)}
(\textbf{\emph{p}}_2,\textbf{\emph{p}}_1)
-N_0^2 |u_0(\textbf{\emph{p}}_1)|^2 |u_0(\textbf{\emph{p}}_2)|^2 .
\end{eqnarray}
Then, the two-pion correlation function is
\begin{eqnarray}
\label{Cp1p2_2}
C_2{(\textbf{\emph{p}}_1,\textbf{\emph{p}}_2)}=1 +\frac{|G^{(1)}(\textbf{\emph{p}}_1,\textbf{\emph{p}}_2)|^2
-N_0^2|u_0(\textbf{\emph{p}}_1)|^2 |u_0(\textbf{\emph{p}}_2)|^2} {G^{(1)}(\textbf{\emph{p}}_1,\textbf{\emph{p}}_1)\,
G^{(1)}(\textbf{\emph{p}}_2,\textbf{\emph{p}}_2)}.
\end{eqnarray}
In the nearly completely coherent case with almost all particles in the ground
condensate state, $N_0\to N$, the two terms in the numerator approximately cancel
each other and the correlation function approaches 1.  For the other extreme of
a completely chaotic source with $N_0<<N$, the second term in the numerator can
be neglected, and we have
\begin{equation}
\label{Cp1p2_3}
C_2{(\textbf{\emph{p}}_1,\textbf{\emph{p}}_2)}=1 +\frac{|G^{(1)}(\textbf{\emph{p}}_1,
\textbf{\emph{p}}_2)|^2}{G^{(1)}(\textbf{\emph{p}}_1,\textbf{\emph{p}}_1)\, G^{(1)}
(\textbf{\emph{p}}_2,\textbf{\emph{p}}_2)}.
\end{equation}

\begin{figure}[htb]
\includegraphics[width=0.85\columnwidth]{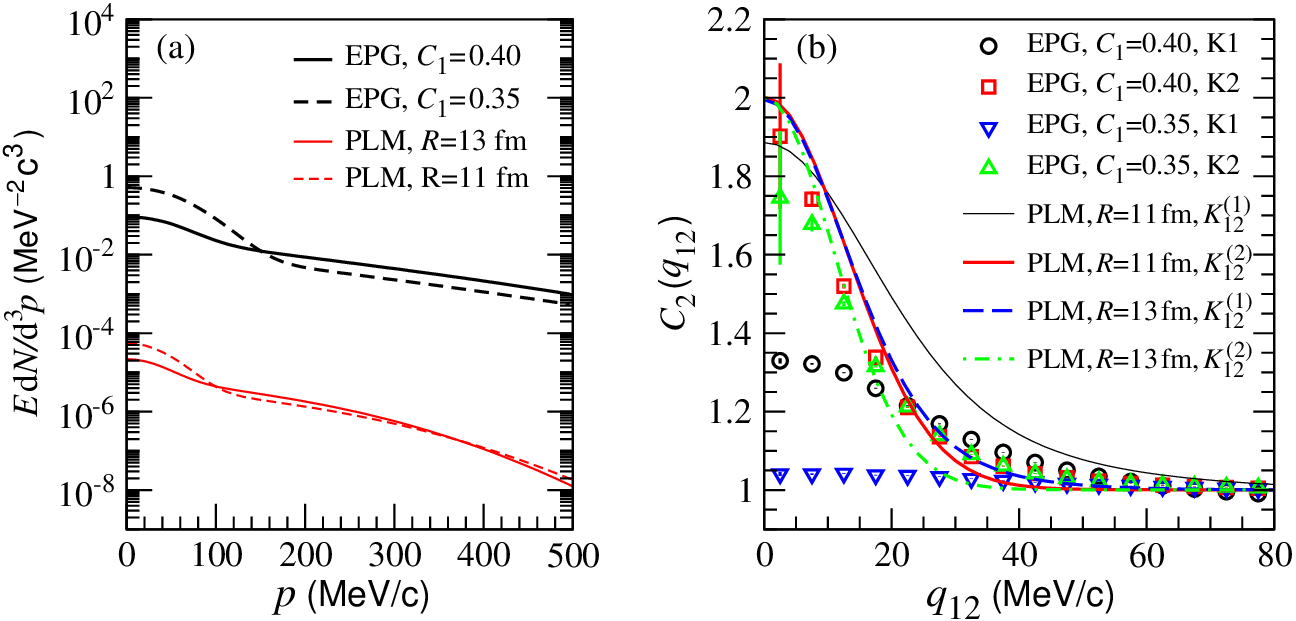}
\caption{(Color online) The invariant single-pion momentum distributions (a) and the
two-pion correlation functions (b) in the EPG model and PLM \cite{Pratt93,CsorgoZimanyi97}. }
\label{zfspC2}
\end{figure}

In Fig.~\ref{zfspC2}(a), the thick solid and dashed curves show the invariant
single-pion momentum distributions in the EPG model with the parameters $C_1=$
0.40 and 0.35, respectively.  The particle number $N$ and temperature $T$ are
taken to be 1200 and 100 MeV.  It can be seen that the momentum distribution for
$C_1=0.35$ has a more obvious enhancement in low momentum region compared to that
for $C_1=0.40$.  It is because the source size for $C_1=0.35$ is small than that
for $C_1=0.40$, and the higher condensation for the smaller source leads to more
pions condensed in the ground state and with small momenta.  The thin solid and
dashed curves in Fig.~\ref{zfspC2}(a) represent the exclusive invariant
single-pion momentum distributions, calculated with the formulas in Ref.
\cite{CsorgoZimanyi97}, in the PLM for identical particle number $N=1200$ and
with the source radii $R=$ 11 and 13 fm, respectively.  The other parameters
in the calculations are taken to be $\sigma_x=2$ fm and $T=120$ MeV as in Ref.
\cite{CsorgoZimanyi97}.
It also can be seen that the momentum distribution for the smaller source has a
more obvious enhancement in low momentum region than that for the larger source.
Because the PLM calculation formulas are non-relativistic \cite{CsorgoZimanyi97},
the invariant momentum distributions for the PLM sources decrease more rapidly in
high momentum region than those for the EPG sources.

In Fig.~\ref{zfspC2}(b) we show the two-pion correlation functions for the EPG and PLM
sources as in Fig.~\ref{zfspC2}(a).  Here, $q_{12}$ is invariant relative momentum of
the two pions, $q_{12}=\sqrt{-(p_1-p_2)^{\mu}(p_1-p_2)_{\mu}}$.
For the EPG sources, $K1$ and $K2$ denote the results calculated in the momentum
intervals $|\textbf{\emph{p}}_1+\textbf{\emph{p}}_2|/2<150$ MeV/$c$ and $|\textbf{\emph{p}}_1
+\textbf{\emph{p}}_2|/2>150$ MeV/$c$, respectively.  The results for the PLM sources are
calculated with the formulas in Ref. \cite{CsorgoZimanyi97}.  Here, $K_{12}^{(1)}$ and
$K_{12}^{(2)}$ denote the results calculated for $|\textbf{\emph{p}}_1+\textbf{\emph{p}}_2|
/2=100$ MeV/$c$ and $|\textbf{\emph{p}}_1+\textbf{\emph{p}}_2|/2=250$ MeV/$c$, respectively.
For the EPG sources, the intercepts of two-pion correlation functions decrease with decreasing
source size because the condensation is significant in small system.  Also, the intercepts of
two-pion correlation functions calculated in the lower momentum interval are smaller than those
calculated in the higher momentum interval because of the condensation.  For the PLM sources,
the intercept of two-pion correlation function approaches to two \cite{CsorgoZimanyi97},
except for the result in the case of the small radius and momentum.

\subsection{Calculations of multi-pion correlation functions in EPG model}

Generalizing Eq. (\ref{Cp1p2_1}), the three- and four-pion correlation functions
are defined as,
\begin{equation}
\label{Cp1p2p3_1}
C_3(\textbf{\emph{p}}_1,\textbf{\emph{p}}_2,\textbf{\emph{p}}_3)=\frac{G^{(3)}
(\textbf{\emph{p}}_1,\textbf{\emph{p}}_2,\textbf{\emph{p}}_3;\textbf{\emph{p}}_1,
\textbf{\emph{p}}_2,\textbf{\emph{p}}_3)}{G^{(1)}(\textbf{\emph{p}}_1;
\textbf{\emph{p}}_1)\,G^{(1)}(\textbf{\emph{p}}_2;\textbf{\emph{p}}_2)\,
G^{(1)}(\textbf{\emph{p}}_3;\textbf{\emph{p}}_3)},
\end{equation}
\begin{equation}
\label{Cp1p2p3p4_1}
C_4(\textbf{\emph{p}}_1,\textbf{\emph{p}}_2,\textbf{\emph{p}}_3,\textbf{\emph{p}}_4)
=\frac{G^{(4)}(\textbf{\emph{p}}_1,\textbf{\emph{p}}_2,\textbf{\emph{p}}_3
\textbf{\emph{ p}}_4;\textbf{\emph{p}}_1,\textbf{\emph{p}}_2,\textbf{\emph{p}}_3,
\textbf{\emph{ p}}_4)}{G^{(1)}(\textbf{\emph{p}}_1;\textbf{\emph{p}}_1)\,
G^{(1)}(\textbf{\emph{p}}_2;\textbf{\emph{p}}_2)\,G^{(1)}(\textbf{\emph{p}}_3;
\textbf{\emph{p}}_3)\,G^{(1)}(\textbf{\emph{p}}_4;\textbf{\emph{p}}_4)},
\end{equation}
where
\begin{eqnarray}
G^{(n)}(\textbf{\emph{p}}_1,\dots,\textbf{\emph{p}}_n;\textbf{\emph{p}}_1,
\dots,\textbf{\emph{p}}_n)=\sum_{k_1,\dots,k_n,l_1,\dots,l_n}\!\!\!\!
&&u_{k_1}^*(\textbf{\emph{p}}_1)\cdots u_{k_n}^*(\textbf{\emph{p}}_n)
u_{l_1}(\textbf{\emph{p}}_1)\cdots u_{l_n}(\textbf{\emph{p}}_n)\nonumber\\
&&\times \langle {\hat a}^{\dag}_{k_1}\cdots{\hat a}^{\dag}_{k_n}
{\hat a}_{l_1}\cdots{\hat a}_{l_n}\rangle
\end{eqnarray}
is the $n$-particle density matrix in momentum space.

For the EPG source with Bose-Einstein condensation, the multi-pion correlation
functions can be written as,
\begin{equation}
\label{Cp1p2p3_2}
C_3(\textbf{\emph{p}}_1,\textbf{\emph{p}}_2,\textbf{\emph{p}}_3)=1+R(1,2)+R(1,3)
+R(2,3)+R(1,2,3),
\end{equation}
\begin{eqnarray}
\label{Cp1p2p3p4_2}
&&\hspace{-8mm}C_4(\textbf{\emph{p}}_1,\textbf{\emph{p}}_2,\textbf{\emph{p}}_3,
\textbf{\emph{p}}_4)=1+R(1,2)+R(1,3)+R(1,4)+R(2,3)+R(2,4)+R(3,4)
\nonumber\\
&&\hspace*{32mm}+R(1,2,3)+R(1,2,4)+R(1,3,4)+R(2,3,4)
\nonumber\\
&&\hspace*{32mm}+R(1,2)R(3,4)+R(1,3)R(2,4)+R(1,4)R(2,3)
\nonumber\\
&&\hspace*{32mm}+R(1,2,3,4)+R(1,2,4,3)+R(1,3,2,4),
\end{eqnarray}
where
\begin{equation}
\label{R12_1}
R(i,j)=
\frac{|G^{(1)}(\textbf{\emph{p}}_i,\textbf{\emph{p}}_j)|^2
-N_0^2|u_0(\textbf{\emph{p}}_i)|^2 |u_0(\textbf{\emph{p}}_j)|^2} {G^{(1)}(\textbf{\emph{p}}_i,\textbf{\emph{p}}_i)\,
G^{(1)}(\textbf{\emph{p}}_j,\textbf{\emph{p}}_j)}
\end{equation}
\begin{eqnarray}
\label{R123_1}
&&R(i,j,k)=\nonumber\\
&&\hspace*{5mm}2\frac{{\rm Re}\big[G^{(1)}(\textbf{\emph{p}}_i,\textbf{\emph{p}}_j)
G^{(1)}(\textbf{\emph{p}}_j,\textbf{\emph{p}}_k)G^{(1)}(\textbf{\emph{p}}_k,
\textbf{\emph{p}}_i) -N_0^3 f_3(\textbf{\emph{p}}_i,\textbf{\emph{p}}_j,
\textbf{\emph{p}}_k)\big]}{G^{(1)}(\textbf{\emph{p}}_i,\textbf{\emph{p}}_i) G^{(1)}(\textbf{\emph{p}}_j,\textbf{\emph{p}}_j)G^{(1)}(\textbf{\emph{p}}_k,
\textbf{\emph{p}}_k)},~~~~
\end{eqnarray}
\begin{eqnarray}
\label{R1234_1}
&&R(i,j,k,l)=\nonumber\\
&&\hspace*{5mm}2\frac{{\rm Re}\big[G^{(1)}\!(\textbf{\emph{p}}_{\!i},
\textbf{\emph{p}}_{\!j})G^{(1)}\!(\textbf{\emph{p}}_{\!j},\textbf{\emph{p}}_{\!k})
G^{(1)}\!(\textbf{\emph{p}}_{\!k},\textbf{\emph{p}}_{\!l})
G^{(1)}\!(\textbf{\emph{p}}_{\!l},\textbf{\emph{p}}_{\!i})
-N_0^4 f_4(\textbf{\emph{p}}_{\!i},\textbf{\emph{p}}_{\!j},
\textbf{\emph{p}}_{\!k},\textbf{\emph{p}}_{\!l})\big]}
{G^{(1)}(\textbf{\emph{p}}_i,\textbf{\emph{p}}_i)G^{(1)}(\textbf{\emph{p}}_j,
\textbf{\emph{p}}_j)G^{(1)}(\textbf{\emph{p}}_k,\textbf{\emph{p}}_k) G^{(1)}(\textbf{\emph{p}}_l,\textbf{\emph{p}}_l)}.
\end{eqnarray}
Here, $R(i,j)$, $[R(i,j)R(k,l)]$, $R(i,j,k)$, and $R(i,j,k,l)$ denote the correlations
of single pion pair, double pion pair, pure pion-triplet interference or true three-pion
correlator \cite{Liu86,HeiZhaSug}, and pure pion-quadruplet interference, respectively.
The functions $f_3(\textbf{\emph{p}}_i,\textbf{\emph{p}}_j,\textbf{\emph{p}}_k)$ and $f_4(\textbf{\emph{p}}_i,\textbf{\emph{p}}_j,\textbf{\emph{p}}_k,
\textbf{\emph{p}}_l)$ in Eqs. (\ref{R123_1}) and (\ref{R1234_1}) are given by
\begin{eqnarray}
f_3(\textbf{\emph{p}}_i,\textbf{\emph{p}}_j,\textbf{\emph{p}}_k)
&=&G^{(1)}(\textbf{\emph{p}}_i,\textbf{\emph{p}}_j)u_0(\textbf{\emph{p}}_i)
u_0^*(\textbf{\emph{p}}_j)|u_0(\textbf{\emph{p}}_k)|^2/{N_0}\nonumber\\
&+&G^{(1)}(\textbf{\emph{p}}_j,\textbf{\emph{p}}_k)u_0(\textbf{\emph{p}}_j)
u_0^*(\textbf{\emph{p}}_k)|u_0(\textbf{\emph{p}}_i)|^2/{N_0}\nonumber\\
&+&G^{(1)}(\textbf{\emph{p}}_k,\textbf{\emph{p}}_i)u_0(\textbf{\emph{p}}_k)
u_0^*(\textbf{\emph{p}}_i)|u_0(\textbf{\emph{p}}_j)|^2/{N_0}\nonumber\\
&-&2|u_0(\textbf{\emph{p}}_i)|^2|u_0(\textbf{\emph{p}}_j)|^2|u_0(\textbf{\emph{p}}_k)|^2,
\end{eqnarray}
\begin{eqnarray}
f_4(\textbf{\emph{p}}_i,\textbf{\emph{p}}_j,\textbf{\emph{p}}_k,\textbf{\emph{p}}_l)
&=&G^{(1)}(\textbf{\emph{p}}_i,\textbf{\emph{p}}_j)G^{(1)}(\textbf{\emph{p}}_j,
\textbf{\emph{p}}_k)u_0(\textbf{\emph{p}}_i)u_0^*(\textbf{\emph{p}}_k)
|u_0(\textbf{\emph{p}}_l)|^2/{N_0^2}\nonumber\\
&+&G^{(1)}(\textbf{\emph{p}}_i,\textbf{\emph{p}}_j)G^{(1)}(\textbf{\emph{p}}_l,
\textbf{\emph{p}}_i)u_0^*(\textbf{\emph{p}}_j)u_0(\textbf{\emph{p}}_l)
|u_0(\textbf{\emph{p}}_k)|^2/{N_0^2}\nonumber\\
&+&G^{(1)}(\textbf{\emph{p}}_j,\textbf{\emph{p}}_k)G^{(1)}(\textbf{\emph{p}}_k,
\textbf{\emph{p}}_l)u_0(\textbf{\emph{p}}_j)u_0^*(\textbf{\emph{p}}_l)
|u_0(\textbf{\emph{p}}_i)|^2/{N_0^2}\nonumber\\
&+&G^{(1)}(\textbf{\emph{p}}_l,\textbf{\emph{p}}_i)G^{(1)}(\textbf{\emph{p}}_k,
\textbf{\emph{p}}_l)u_0^*(\textbf{\emph{p}}_i)u_0(\textbf{\emph{p}}_k)
|u_0(\textbf{\emph{p}}_j)|^2/{N_0^2}\nonumber\\
&+&G^{(1)}(\textbf{\emph{p}}_l,\textbf{\emph{p}}_i)G^{(1)}(\textbf{\emph{p}}_j,
\textbf{\emph{p}}_k)u_0^*(\textbf{\emph{p}}_i)u_0(\textbf{\emph{p}}_j)
u_0^*(\textbf{\emph{p}}_k)u_0(\textbf{\emph{p}}_l)/{N_0^2}\nonumber\\
&+&G^{(1)}(\textbf{\emph{p}}_i,\textbf{\emph{p}}_j)G^{(1)}(\textbf{\emph{p}}_k,
\textbf{\emph{p}}_l)u_0(\textbf{\emph{p}}_i)u_0^*(\textbf{\emph{p}}_j)
u_0(\textbf{\emph{p}}_k)u_0^*(\textbf{\emph{p}}_l)/{N_0^2}\nonumber\\
&-&2G^{(1)}(\textbf{\emph{p}}_i,\textbf{\emph{p}}_j)u_0(\textbf{\emph{p}}_i)
u_0^*(\textbf{\emph{p}}_j)|u_0(\textbf{\emph{p}}_k)u_0(\textbf{\emph{p}}_l)
|^2/{N_0}\nonumber\\
&-&2G^{(1)}(\textbf{\emph{p}}_j,\textbf{\emph{p}}_k)u_0(\textbf{\emph{p}}_j)
u_0^*(\textbf{\emph{p}}_k)|u_0(\textbf{\emph{p}}_i)u_0(\textbf{\emph{p}}_l)
|^2/{N_0}\nonumber\\
&-&2G^{(1)}(\textbf{\emph{p}}_k,\textbf{\emph{p}}_l)u_0(\textbf{\emph{p}}_k)
u_0^*(\textbf{\emph{p}}_l)|u_0(\textbf{\emph{p}}_i)u_0(\textbf{\emph{p}}_j)
|^2/{N_0}\nonumber\\
&-&2G^{(1)}(\textbf{\emph{p}}_l,\textbf{\emph{p}}_i)u_0(\textbf{\emph{p}}_l)
u_0^*(\textbf{\emph{p}}_i)|u_0(\textbf{\emph{p}}_j)u_0(\textbf{\emph{p}}_k)
|^2/{N_0}\nonumber\\
&+&3|u_0(\textbf{\emph{p}}_i)|^2|u_0(\textbf{\emph{p}}_j)|^2|u_0(\textbf{\emph{p}}_k)|^2
|u_0(\textbf{\emph{p}}_l)|^2.
\end{eqnarray}
In the nearly completely coherent case, almost all particles are in the ground
condensate state, functions $f_3(\textbf{\emph{p}}_i, \textbf{\emph{p}}_j,
\textbf{\emph{p}}_k) \to |u_0(\textbf{\emph{p}}_i)|^2 |u_0(\textbf{\emph{p}}_j)|^2| u_0(\textbf{\emph{p}}_k)|^2$ and $f_4(\textbf{\emph{p}}_i, \textbf{\emph{p}}_j,
\textbf{\emph{p}}_k, \textbf{\emph{p}}_l) \to |u_0(\textbf{\emph{p}}_i)|^2
|u_0(\textbf{\emph{p}}_j)|^2| u_0(\textbf{\emph{p}}_k)|^2 u_0(\textbf{\emph{p}}_l)|^2$,
and the two terms in the numerators in Eqs. (\ref{R12_1}), (\ref{R123_1}) and
(\ref{R1234_1}) cancel each other approximately.  So, the two-pion, three-pion, and
four-pion correlation functions approaches 1 in the completely coherent case.

\begin{figure}[htb]
\vspace*{3mm}
\includegraphics[width=0.65\columnwidth]{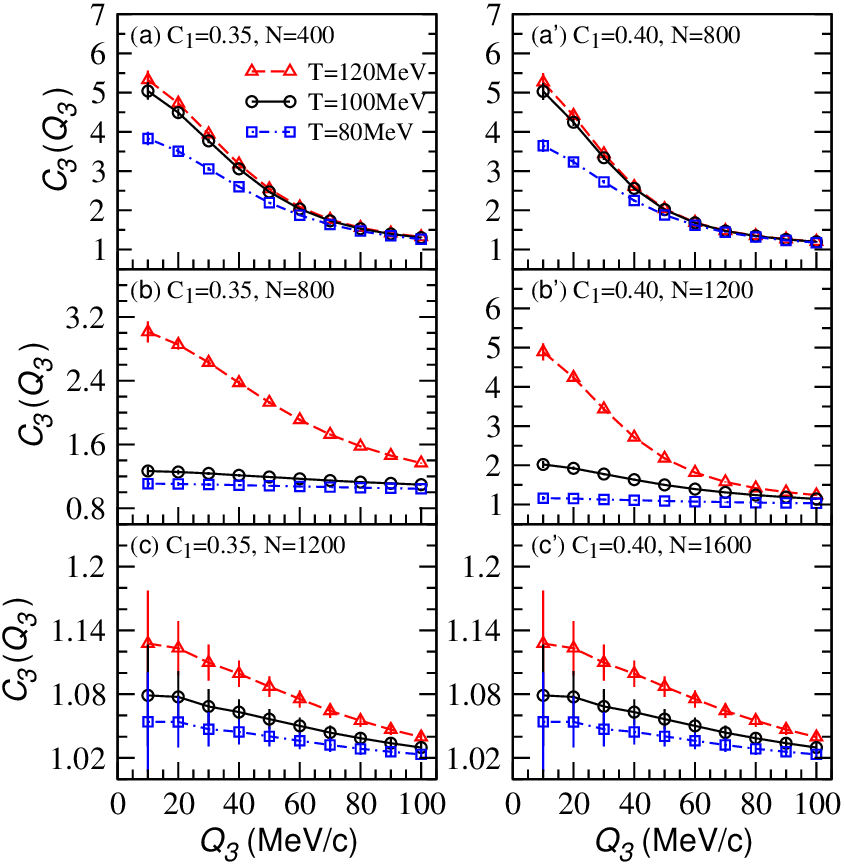}
\caption{(Color online) Three-pion correlation functions for the EPG sources with
different temperatures and particle numbers.}
\label{zf1_z32}
\end{figure}

In EPG model, we can calculate density matrices $G^{(1)}(\textbf{\emph{p}}_i,
\textbf{\emph{p}}_j)$ and wave function $u_0(\textbf{\emph{p}})$ \cite{LiuRuZhangWong14},
and then obtain three- and four-pion correlation functions with Eqs. (\ref{Cp1p2p3_2})
--- (\ref{R1234_1}).  In Fig. \ref{zf1_z32} we plot the three-pion correlations as
a function of $Q_3$ for the EPG sources with different temperatures and particle
numbers.  Here, the Lorentz-invariant momentum of the three pions with four-dimension
momenta $p_i=(E_i,\textbf{\emph{p}}_i)$ ($i=1,2,3$) is defined
as
\begin{equation}
Q_3=\sqrt{q_{12}^2+q_{13}^2+q_{23}^2},
\end{equation}
where
\begin{equation}
q_{ij}=\sqrt{-(p_i-p_j)^{\mu}(p_i-p_j)_{\mu}}.
\end{equation}

The three-pion correlation functions for the sources with small particle numbers
($N=400$ for $C_1=0.35$ and $N=800$ for $C_1=0.40$) are high.  They decrease with
increasing $N$ because the source with large particle number has significant
condensation.  For fixed source particle number $N$, the three-pion correlation
function increases with increasing temperature because the condensation fraction is low
at high temperatures.  For fixed $N$ and $T$, the three-pion correlation functions
for the sources with $C_1=0.35$ are lower than those for the sources with $C_1=0.40$
because the source with a small $C_1$ has small characteristic length and high
condensation fraction.

\begin{figure}[htb]
\vspace*{3mm}
\includegraphics[width=0.65\columnwidth]{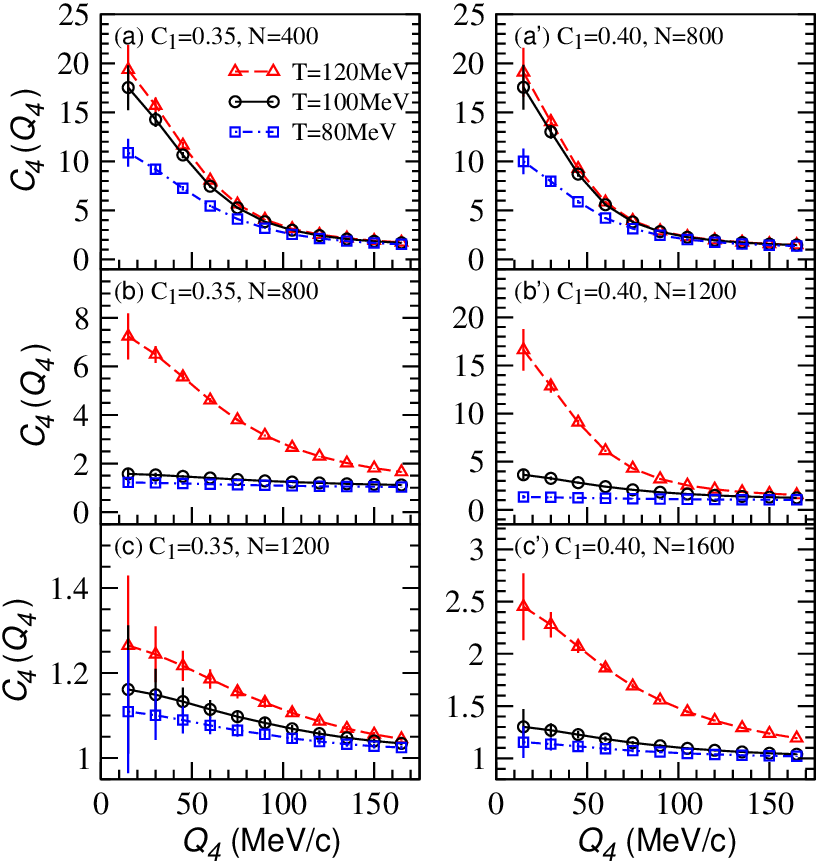}
\caption{(Color online) Four-pion correlation functions for the EPG sources with
different temperatures and particle numbers.}
\label{zf1_z42}
\end{figure}

In Fig.~\ref{zf1_z42}, we plot the four-pion correlations as a function of $Q_4$,
\begin{equation}
Q_4=\sqrt{q_{12}^2+q_{13}^2+q_{23}^2+q_{14}^2+q_{24}^2+q_{34}^2},
\end{equation}
for the EPG sources with different temperatures and the particle numbers.  The
four-pion correlation functions exhibit the similar variations with source
particle number, temperature, and parameter $C_1$ as those of the three-pion
correlation functions.  However, the four-pion correlation functions are higher
than the corresponding three-pion correlation functions because there are more
contributions of the correlations of single pion pair, double pion pair, pure
pion-triplet interference, and the contribution of the correlations of pure
pion-quadruplet interference in four-pion correlation functions.

\section{Analyses of multi-pion correlations in EPG model}
In Ref. \cite{ALICE-PRC16}, the ALICE collaboration measured the three- and four-pion
correlation functions in Pb-Pb collisions at $\sqrt{s_{NN}}=2.76$~TeV, in the average
transverse-momentum intervals $0.16<K_{T3,T4}<0.3$~GeV/$c$ and $0.3<K_{T3,T4}<1$~GeV/$c$,
where
\begin{equation}
K_{T3}=\frac{|\textbf{\emph{p}}_{T1}+\textbf{\emph{p}}_{T2}+\textbf{\emph{p}}_{T3}|}{3},
~~~~~
K_{T4}=\frac{|\textbf{\emph{p}}_{T1}+\textbf{\emph{p}}_{T2}+\textbf{\emph{p}}_{T3}+
\textbf{\emph{p}}_{T4}|}{4}.
\end{equation}
In this section we shall investigate the three- and four-pion correlation functions
in the EPG model in different transverse-momentum intervals in order to compare the
model results with experimental data.

\subsection{Three-pion correlations in EPG model}
In the EPG model considered, the average momentum of the particles emitted from the ground
state (coherent emission) is smaller than that of the particles emitted from the excited
states (chaotic emission).  So, the multi-pion correlation functions for the EPG source
with a finite condensation fraction are momentum dependent.

\begin{figure}[htb]
\vspace*{3mm}
\includegraphics[width=0.85\columnwidth]{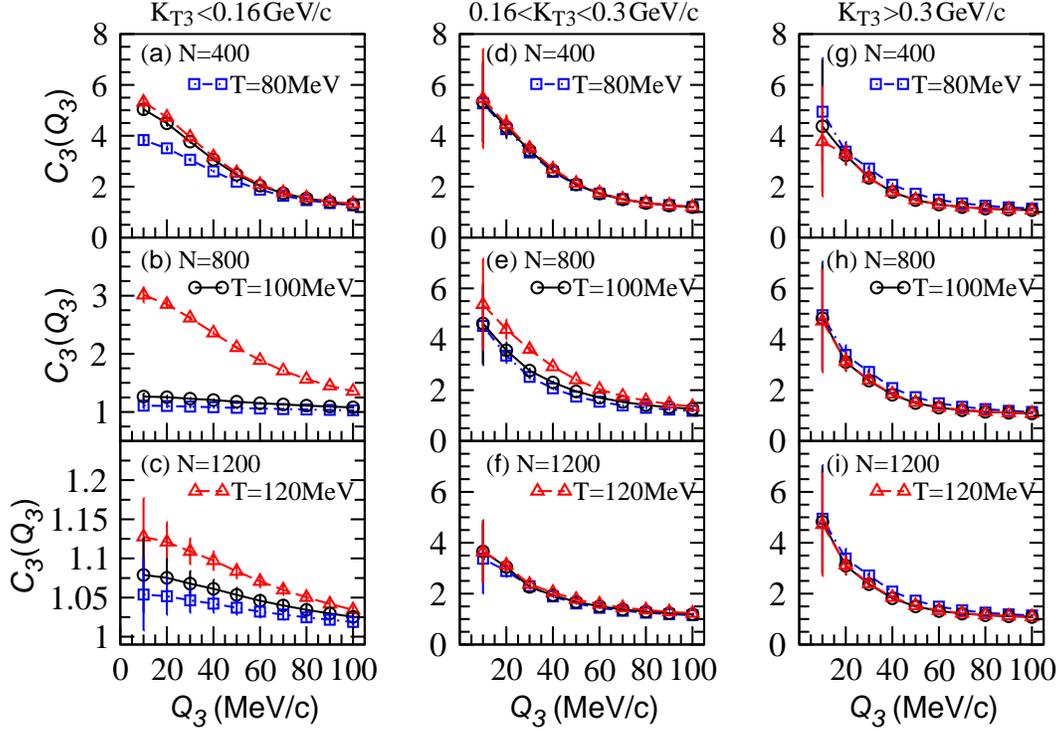}
\caption{(Color online) Three-pion correlation functions for the EPG sources
with $C_1=0.35$ and different particle numbers, in the transverse-momentum intervals
$K_{T3}<0.16$~GeV/$c$ [(a)--(c)], $0.16<K_{T3}<0.3$~GeV/$c$ [(d)--(f)], and
$K_{T3}>0.3$~GeV/$c$ [(g)--(i)]. }
\label{zf2-3p-35}
\end{figure}

We plot in Fig. \ref{zf2-3p-35} the three-pion correlation functions for the EPG
sources with $C_1=0.35$ and the particle numbers $N=400$, 800, and 1200.
The transverse-momentum cuts $K_{T3}<0.16$~GeV/$c$, $0.16<K_{T3}<0.3$~GeV/$c$, and
$K_{T3}>0.3$~GeV/$c$ are applied in the simulated calculations of the correlation
functions shown in Figs.~\ref{zf2-3p-35}(a)--(c), Figs.~\ref{zf2-3p-35}(d)--(f),
and Figs.~\ref{zf2-3p-35}(g)--(i), respectively.
In the lowest momentum interval $K_{T3}<0.16$~GeV/$c$ [Figs.~\ref{zf2-3p-35}(a)--(c)],
the correlation functions increase with source temperature $T$ and decrease with
increasing particle number $N$ in the source.  The reasons are that the source has a
lower condensation fraction at higher temperature than that at lower temperature, and
the condensation fraction increases with increasing particle number in the source.
For $N=400$, the results in Fig.~\ref{zf2-3p-35}(a) show that the intercepts of the
three-pion correlation functions for the sources with the temperatures higher than
80~MeV approach the maximum 6 when being extrapolated to $Q_3=0$.  This indicates
that the sources with the higher temperatures are almost completely chaotic.
The result of the three-pion correlation function for $T=80$~MeV shown in
Fig.~\ref{zf2-3p-35}(a) indicates that there is a finite fraction of coherent
emission when the source has a temperature of $T=80$~MeV and particle number $N=400$.
However, the results in Fig.~\ref{zf2-3p-35}(c) indicate that all the sources with
the three temperatures have high condensation fractions when $N=1200$.
On the other hand, in the highest momentum interval $K_{T3}>0.3$~GeV/$c$
[Figs.~\ref{zf2-3p-35}(g)--(i)], the high intercepts of correlation functions
indicate that most of the pions with high momenta are emitted chaotically from
excited states, even if the sources with high condensation fractions (with
large $N$) \cite{LiuRuZhangWong14}.  The widths of the correlation functions in the
highest momentum interval are narrower than those in the lowest momentum interval
because the source has a wider spatial distribution for the pions emitted from excited
states than that from ground state \cite{LiuRuZhangWong14}.  The correlation functions for
the sources with $T=80$~MeV are slightly higher than those for the sources with
the higher temperatures in the highest momentum interval because the source spatial
distribution is narrow at low temperature for the chaotic emission from excited states
\cite{LiuRuZhangWong14}.
In the middle momentum interval $0.16<K_{T3}<0.3$~GeV/$c$ [Figs.~\ref{zf2-3p-35}(d)--(f)],
the condensation effect on the correlation functions is weaker than that in the lowest
momentum interval $K_{T3}<0.16$~GeV/$c$, because the number of the pions emitted from
excited states is averagely larger in the middle momentum interval than that in the
lowest momentum interval.  Meanwhile, there is also the influence of source spatial
distributions at different temperatures on the correlation functions in the middle
momentum interval.

\begin{figure}[htb]
\vspace*{3mm}
\includegraphics[width=0.85\columnwidth]{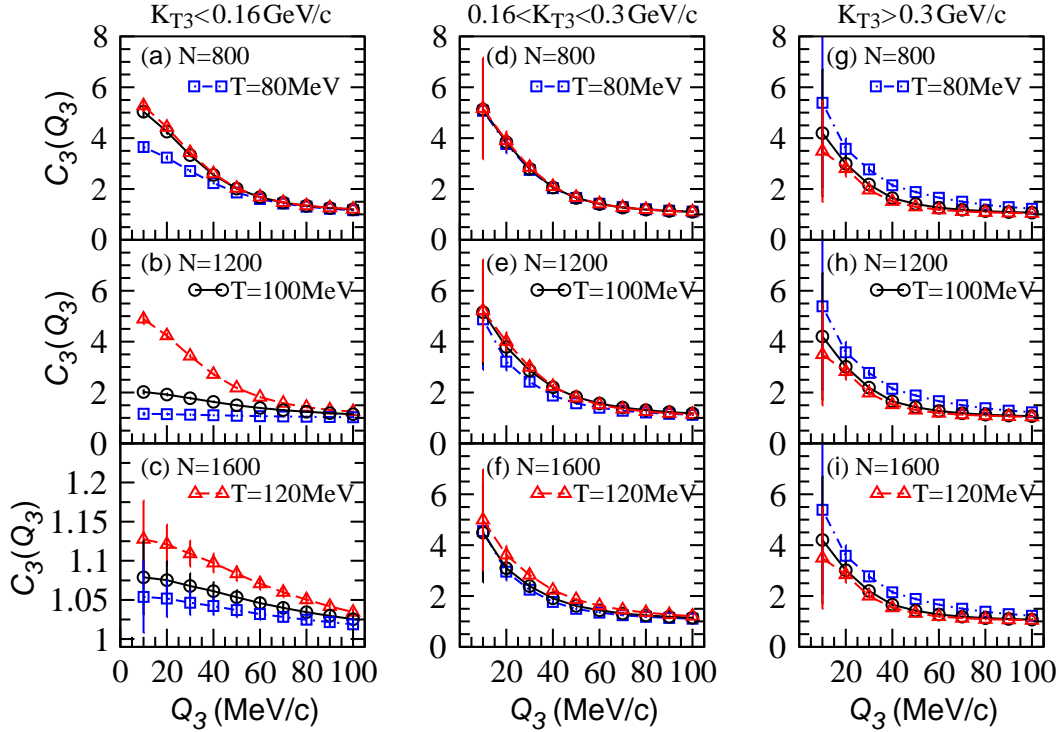}
\caption{(Color online) Three-pion correlation functions for the EPG sources
with different particle numbers and $C_1=0.40$, in the transverse momentum intervals
$K_{T3}<0.16$~GeV/$c$ [(a)--(c)], $0.16<K_{T3}<0.3$~GeV/$c$ [(d)--(f)], and
$K_{T3}>0.3$~GeV/$c$ [(g)--(i)]. }
\label{zf2-3p-40}
\end{figure}

\begin{figure}[htb]
\vspace*{3mm}
\includegraphics[width=0.85\columnwidth]{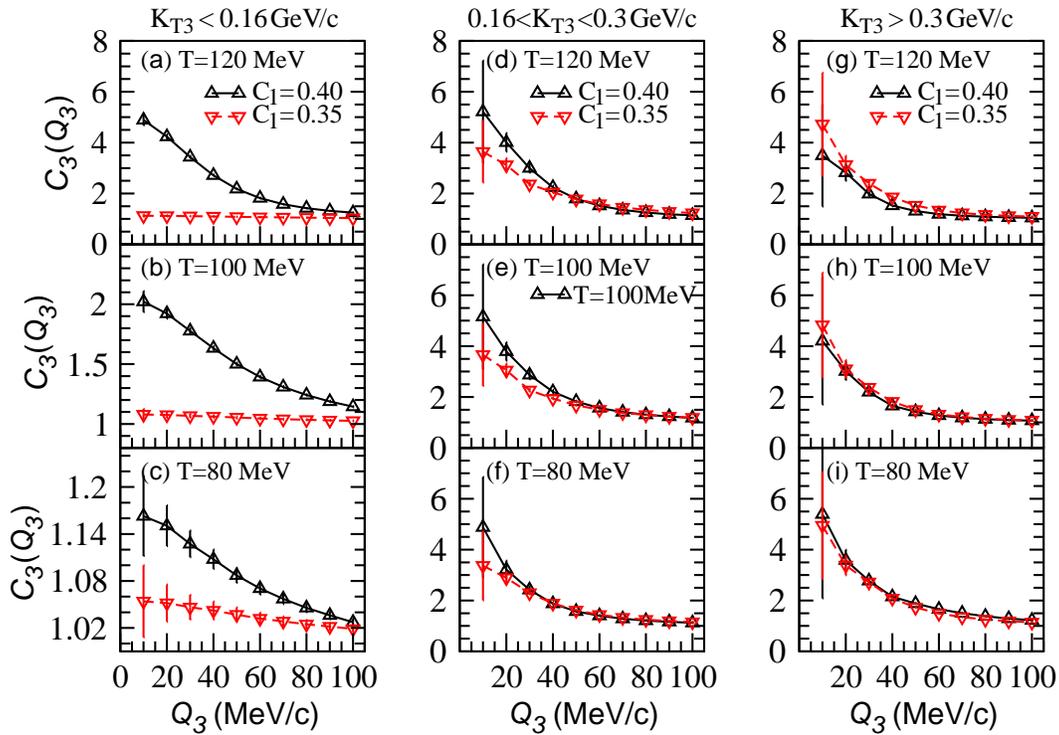}
\caption{(Color online) Three-pion correlation functions for the EPG sources with $C_1=$ 0.35
and 0.40, in the transverse momentum intervals $K_{T3}<0.16$~GeV/$c$ [(a)--(c)], $0.16<K_{T3}
<0.3$~GeV/$c$ [(d)--(f)], and $K_{T3}>0.3$~GeV/$c$ [(g)--(i)].
The particle number is 1200. }
\label{zf2-3p-1200}
\end{figure}

We plot in Fig. \ref{zf2-3p-40} the three-pion correlation functions for the EPG
sources with $C_1=0.40$ and the particle numbers $N=800$, 1200, and 1600.
The transverse-momentum cuts $K_{T3}<0.16$~GeV/$c$, $0.16<K_{T3}<0.3$~GeV/$c$, and
$K_{T3}>0.3$~GeV/$c$ are applied in the simulated calculations of the correlation
functions shown in Figs.~\ref{zf2-3p-35}(a)--(c), Figs.~\ref{zf2-3p-35}(d)--(f), and
Figs.~\ref{zf2-3p-35}(g)--(i), respectively.  One can see that the correlation functions
in Fig. \ref{zf2-3p-40} exhibit the similar variations with source temperature and particle
number in the transverse-momentum intervals as those in Fig. \ref{zf2-3p-35}.  We further
show the comparisons of the three-pion correlation functions for the sources with $C_1=$
0.35 and 0.40 in Fig. \ref{zf2-3p-1200}.  Here, the particle numbers of both the sources
with $C_1=$ 0.35 and 0.40 are 1200.  In the lowest transverse-momentum interval $K_{T3}<
0.16$~GeV/$c$, the three-pion correlation functions for the sources with $C_1=0.35$ are
lower than those for the sources with $C_1=0.40$ at all the temperatures.  It is because
the condensation fraction is high for the source with small $C_1$ and therefore with small
characteristic length $a$.  The differences between the correlation functions for the sources
with $C_1=$ 0.35 and 0.40 become small in the middle transverse-momentum interval
$0.16<K_{T3}<0.3$~GeV/$c$ and almost zero in the highest transverse-momentum interval $K_{T3}>0.3$~GeV/$c$.  This indicates that the condensation effect on the correlation
functions decreases with the increasing average transverse momentum $K_{T3}$ because
the pions emitted chaotically from excited states have high average momentum.

\begin{figure}[htb]
\vspace*{3mm}
\includegraphics[width=0.85\columnwidth]{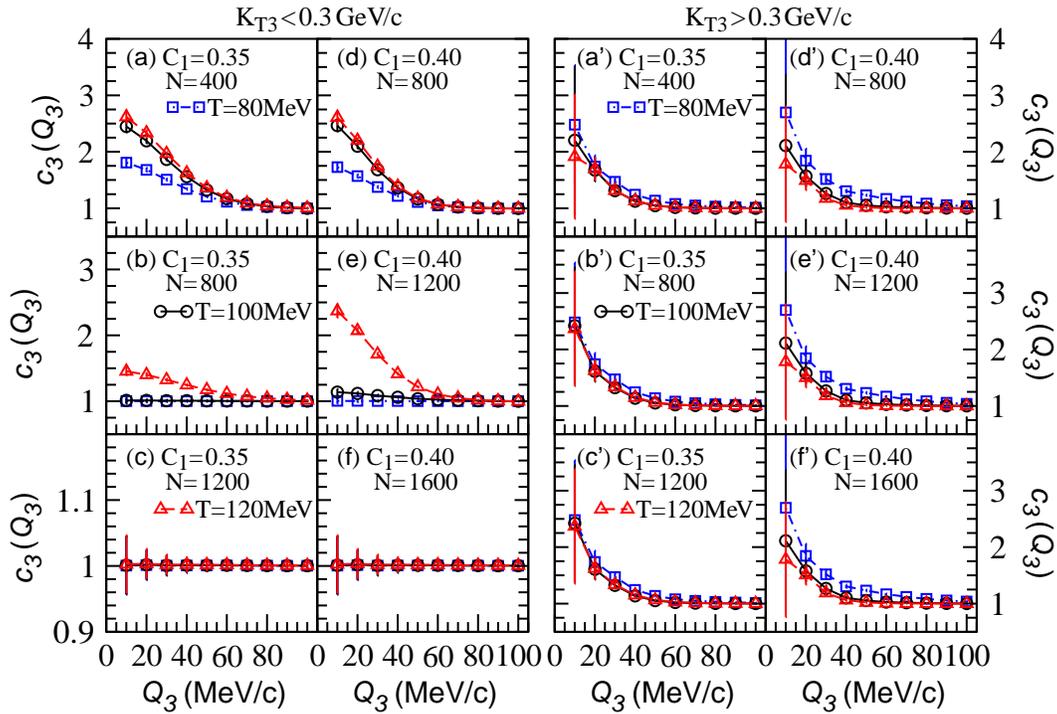}
\caption{(Color online) Three-pion cumulant correlation functions for the EPG sources with
$C_1=$ 0.35 and 0.40, in the transverse momentum intervals $K_{T3}<0.3$~GeV/$c$ and $K_{T3}
>0.3$~GeV/$c$. }
\label{zf2-3p-c3}
\end{figure}

We plot in Fig. \ref{zf2-3p-c3} the three-pion cumulant correlation functions,
$c_3(Q_3)=1+R(1,2,3)$, for the EPG sources with $C_1=$ 0.35 and 0.40 and in the low and high
transverse-momentum intervals $K_{T3}<0.3$~GeV/$c$ and $K_{T3}>0.3$~GeV/$c$.  The particle
numbers of the sources with $C_1=0.35$ are 400, 800, and 1200, and the particle numbers
of the sources with $C_1=0.40$ are 800, 1200, and 1600, respectively.
In the low transverse-momentum interval, $c_3$ decreases with increasing $N$ because
the condensation fraction of source increases with increasing $N$.  As the correlation from
the pure pion-triplet interference, $R(1,2,3)$, approaches zero when any pion pair among the
three pions is uncorrelated \cite{Liu86,HeiZhaSug}, $c_3$ is sensitive to the source
condensation in the low momentum interval.  In the high transverse-momentum interval, $c_3$
is almost independent of the source particle number $N$.  This indicates that most of the
pions with high momenta are emitted chaotically from excited states even if the source
with a considerable condensation fraction (for large $N$).  The correlation function for
$T=80$~MeV is wider than that for the higher temperatures because the source spatial
distribution is narrower at lower temperature than that at higher temperature for the
chaotic emission from excited states \cite{LiuRuZhangWong14}.

\subsection{Four-pion correlations in EPG model}

\begin{figure}[htb]
\includegraphics[width=0.85\columnwidth]{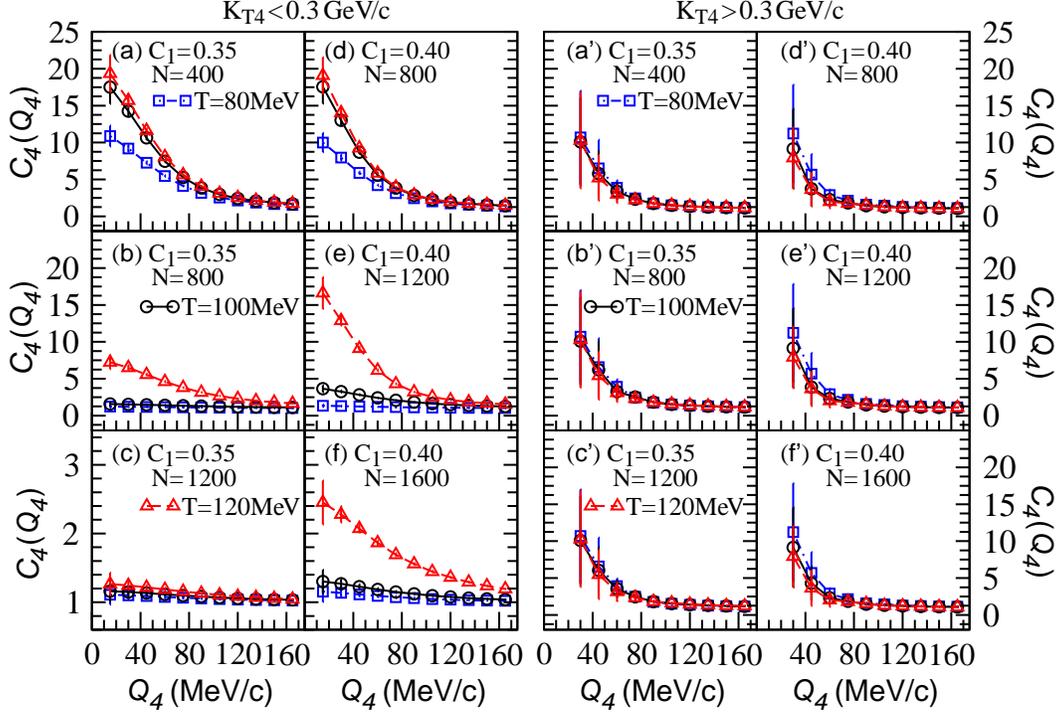}
\caption{(Color online) Four-pion correlation functions $C_4(Q_4)$ for the EPG sources
with $C_1=$ 0.35 and 0.40 in the transverse-momentum intervals $K_{T4}<0.3$~GeV/$c$ and
$K_{T4}>0.3$~GeV/$c$. }
\label{zf2-4p-C4}
\end{figure}

We plot in Fig. \ref{zf2-4p-C4} the four-pion correlation functions for the EPG sources
with $C_1=$ 0.35 and 0.40 and in the low and high transverse-momentum intervals $K_{T4}
<0.3$~GeV/$c$ and $K_{T4}>0.3$~GeV/$c$.  The particle numbers of the sources with $C_1=
0.35$ are 400, 800, and 1200, and the particle numbers of the sources with $C_1=0.40$
are 800, 1200, and 1600, respectively.  In the low transverse-momentum interval, the
results of $C_4(Q_4)$ are sensitive to the source condensation.  They increase with
source temperature $T$ and decrease with increasing particle number $N$ in the source,
because the source has a low condensation fraction at high temperature and the condensation
fraction increases with increasing $N$.  In the high transverse-momentum interval, the
correlation functions have poor statistics in small $Q_4$ bins.  They behave almost
independent of source temperature and particle number.  This indicates that they are
insensitive to the source condensation because most of the pions with high momenta are
emitted chaotically from excited states.  The four-pion correlation functions for the
sources with $T=80$~MeV are slightly higher than those for the sources with the
higher temperatures in the high transverse-momentum interval, as the three-pion correlation
functions behaved.  Because the pions emitted from excited states have wider spatial
distribution than that emitted from ground state \cite{LiuRuZhangWong14}, the widths of the
correlation functions become narrower in the high transverse-momentum interval than those
in the low transverse-momentum interval.

\begin{figure}[htb]
\vspace*{5mm}
\includegraphics[width=0.85\columnwidth]{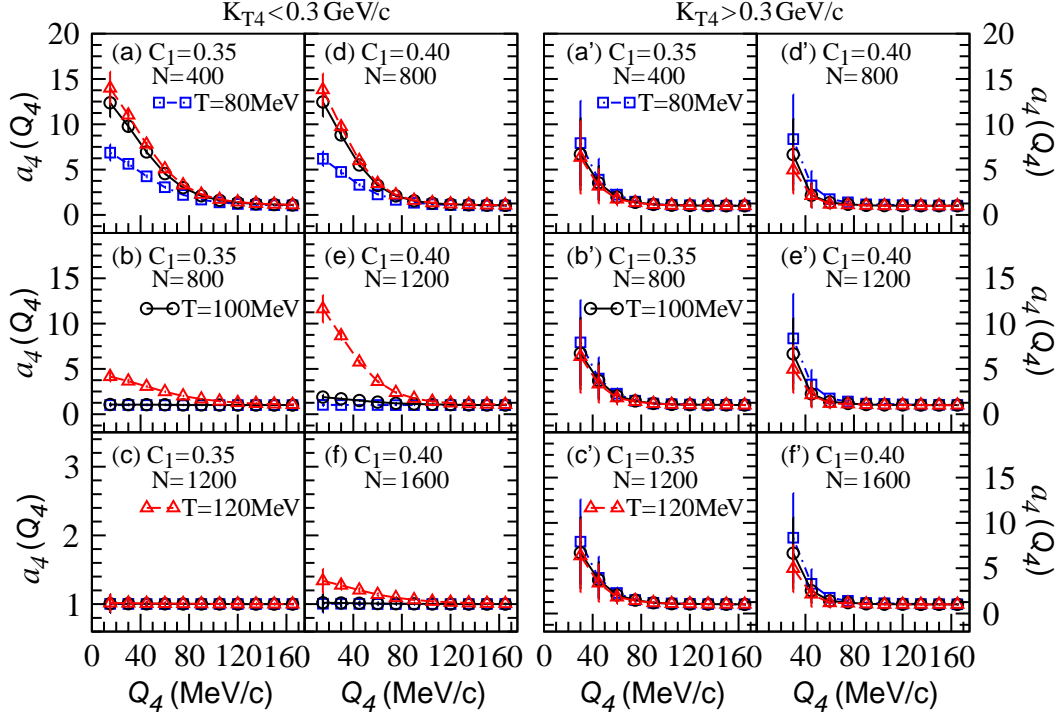}
\caption{(Color online) Four-pion cumulant correlation function $a_4(Q_4)$ for the EPG
sources with $C_1=$ 0.35 and 0.40 in the transverse-momentum intervals $K_{T4}<0.3$~GeV/$c$
and $K_{T4}>0.3$~GeV/$c$. }
\label{zf3-4p-a4}
\end{figure}

In four-pion correlation function $C_4$, there are the contributions of the correlations
of pion pair $R(i,j)$, double pion pair $[R(i,j)R(k,l)]$, pure pion-triplet interference
$R(i,j,k)$, and pure pion-quadruplet interference $R(i,j,k,l)$.  We use $a_4$, $b_4$, and
$c_4$ to denote the four-pion cumulant correlations as \cite{ALICE-PRC16}
\begin{eqnarray}
\!\!\!a_4(\textbf{\emph{p}}_1,\textbf{\emph{p}}_2,\textbf{\emph{p}}_3,\textbf{\emph{p}}_4)=
&&\hspace*{-3mm} 1 + R(1,2,3)+R(1,2,4) + R(1,3,4)+R (2,3,4)\nonumber\\
&&\hspace*{-3mm} + R(1,2,3,4)+ R(1,2,4,3) + R(1,3,2,4)\nonumber\\
&&\hspace*{-3mm} + R(1,2)R(3,4) + R(1,4)R(2,3) + R(1,3)R(2,4);\\
\!\!\!b_4(\textbf{\emph{p}}_1,\textbf{\emph{p}}_2,\textbf{\emph{p}}_3,,\textbf{\emph{p}}_4)=
&&\hspace*{-3mm} 1 + R(1,2,3) + R(1,2,4) + R(1,3,4) + R(2,3,4)\nonumber\\
&&\hspace*{-3mm} + R(1,2,3,4) + R(1,2,4,3) + R(1,3,2,4);
\end{eqnarray}
\begin{eqnarray}
\!\!\!c_4(\textbf{\emph{p}}_1,\textbf{\emph{p}}_2,\textbf{\emph{p}}_3,,\textbf{\emph{p}}_4)=
&&\hspace*{-3mm} 1 + R(1,2,3,4) + R(1,2,4,3) + R(1,3,2,4).
\end{eqnarray}

\begin{figure}[htb]
\vspace*{5mm}
\includegraphics[width=0.85\columnwidth]{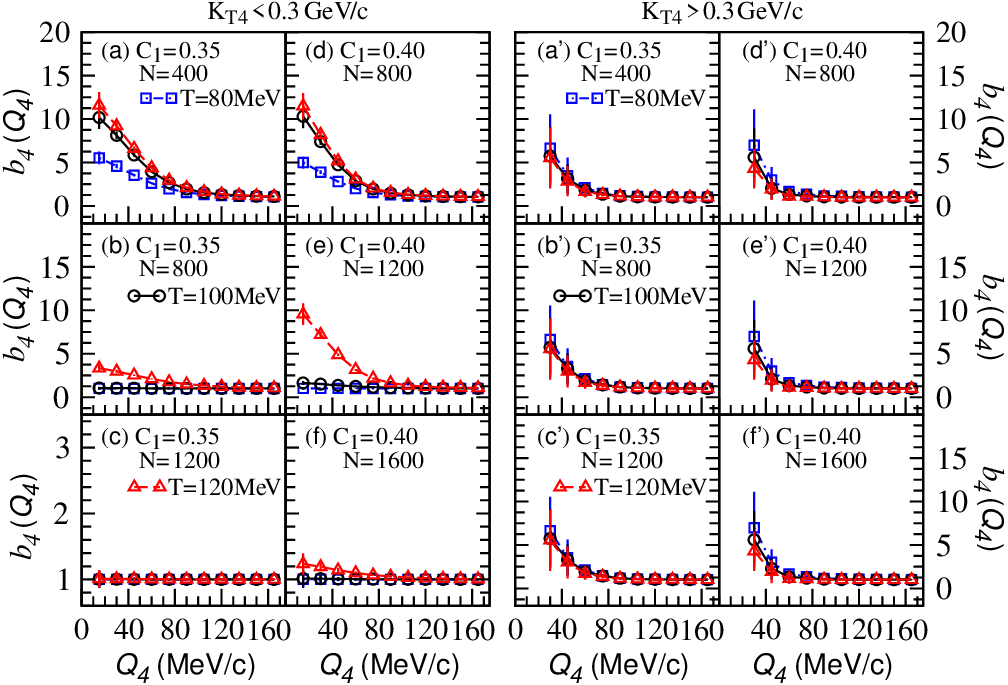}
\caption{(Color online) Four-pion cumulant correlation function $b_4(Q_4)$ for the EPG
sources with $C_1=$ 0.35 and 0.40 in the transverse-momentum intervals $K_{T4}<0.3$~GeV/$c$
and $K_{T4}>0.3$~GeV/$c$. }
\label{zf3-4p-b4}
\end{figure}

\begin{figure}[htb]
\includegraphics[width=0.85\columnwidth]{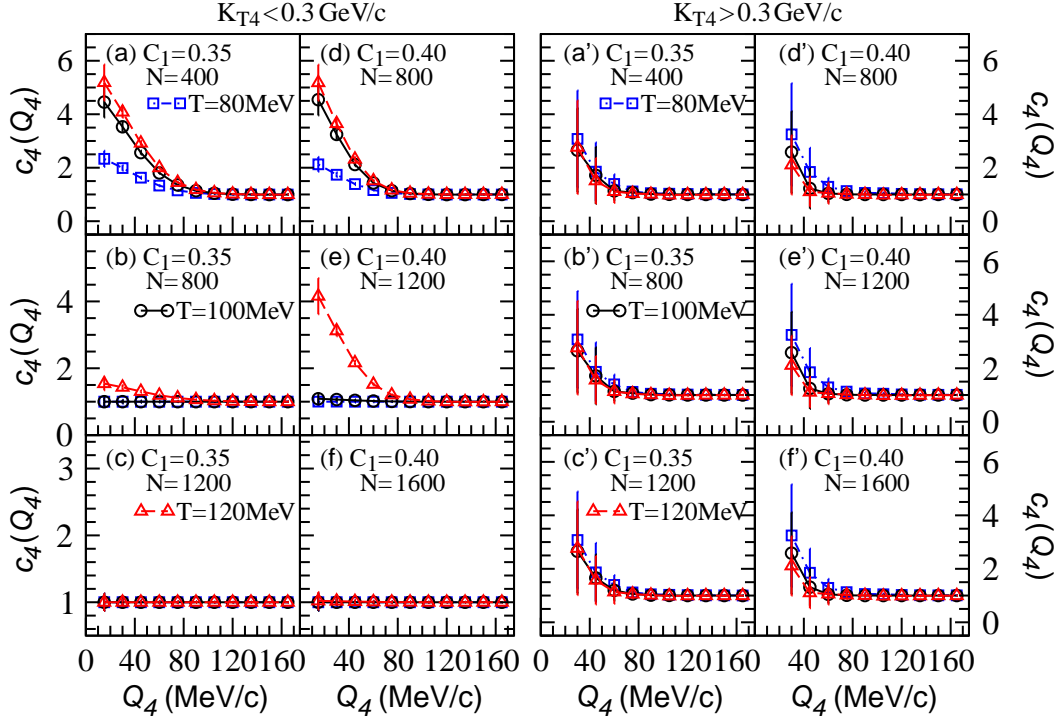}
\caption{(Color online) Four-pion cumulant correlation function $c_4(Q_4)$ for the EPG
sources with $C_1=$ 0.35 and 0.40 in the transverse-momentum intervals $K_{T4}<0.3$~GeV/$c$
and $K_{T4}>0.3$~GeV/$c$. }
\label{zf3-4p-c4}
\end{figure}

We plot in Figs.~\ref{zf3-4p-a4} and \ref{zf3-4p-b4} the four-pion cumulant correlations
$a_4(Q_4)$ and $b_4(Q_4)$ respectively, in the low and high transverse-momentum intervals
$K_{T4}<0.3$~GeV/$c$ and $K_{T4}>0.3$~GeV/$c$.  In $a_4$ the correlations of single pair
are removed.  So, the results of $a_4(Q_4)$ are lower than those of $C_4(Q_4)$
(see Fig.~\ref{zf2-4p-C4}).
In the low transverse-momentum interval, $a_4(Q_4)$ is sensitive to the source condensation
as $C_4(Q_4)$.  It increases with source temperature $T$ and decrease with increasing
particle number $N$ in the source.  However, $a_4(Q_4)$ is also insensitive to source
condensation in the high transverse-momentum interval as $C_4(Q_4)$.
In $b_4$, the correlations of single and double pair are removed.  One can
see from Figs. \ref{zf3-4p-a4} and \ref{zf3-4p-b4} that $b_4(Q_4)$ is slightly lower
than $a_4(Q_4)$ and they have the similar variations with source temperature and particle
number in the low and high transverse-momentum intervals.

We plot in Fig. \ref{zf3-4p-c4} the four-pion cumulant correlation $c_4(Q_4)$ for the EPG
sources with $C_1=$ 0.35 and 0.40 and in the low and high transverse-momentum intervals
$K_{T4}<0.3$~GeV/$c$ and $K_{T4}>0.3$~GeV/$c$.
As $c_4$ contains only the correlations of pure pion-quadruplet interferences, $c_4
(Q_4)$ results are lower than those of $a_4(Q_4)$ and $b_4(Q_4)$.
In the low transverse-momentum interval, $c_4(Q_4)$ decreases with increasing $N$ rapidly.
It drops to 1 for all the temperatures when $N=1200$ for $C_1=0.35$ and $N=1600$ for
$C_1=0.40$ [see Figs. \ref{zf3-4p-c4}(c) and Fig. \ref{zf3-4p-c4}(f)].  In the low
transverse-momentum interval, $c_4(Q_4)$ is more sensitive to source condensation
compared to $a_4(Q_4)$ and $b_4(Q_4)$.

\section{Comparison with experimental data}
In Ref. \cite{ALICE-PRC16}, the ALICE collaboration analyzed the three- and four-pion
correlation functions in Pb-Pb collisions at $\sqrt{s_{NN}}=2.76$~TeV.  They observed
a significant and centrality-independent suppression of the three- and four-pion
correlations.  In this section we shall compare the calculated three- and four-pion
correlation functions in the EPG model with the experimental data for Pb-Pb collisions
at $\sqrt{s_{NN}}=2.76$~TeV \cite{ALICE-PRC16}, and further obtain the information of
source condensation fraction.

\subsection{Three-pion correlations}

\begin{figure}[htb]
\includegraphics[width=0.49\columnwidth]{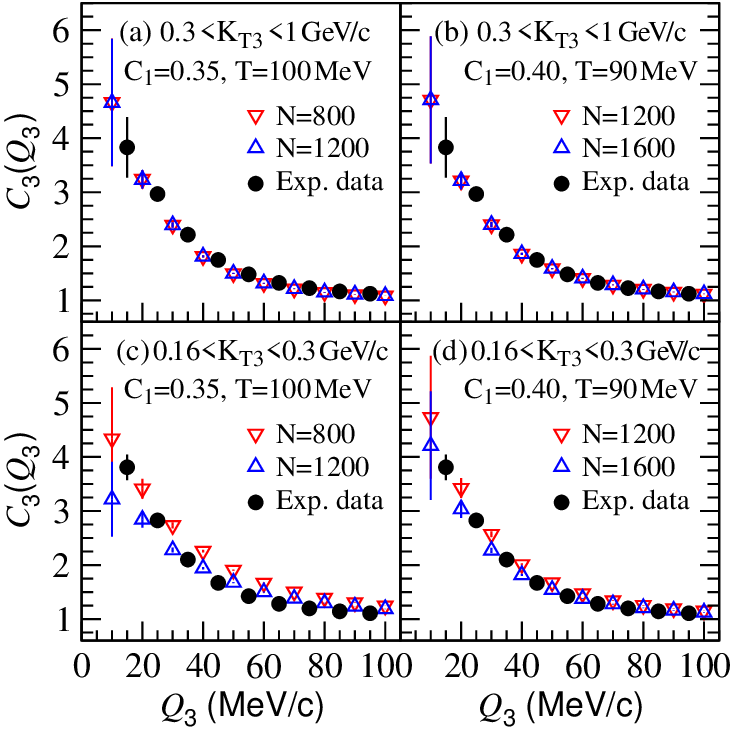}
\vspace*{-3mm}
\caption{(Color online) Comparison of the three-pion correlation function $C_3(Q_3)$
for the EPG sources and the experimental data for Pb-Pb collisions at $\sqrt{s_{NN}}=
2.76$~TeV \cite{ALICE-PRC16}. }
\label{zf4-3p-C3exp}
\end{figure}

In Fig. \ref{zf4-3p-C3exp} we show the comparison of the three-pion correlation functions
$C_3(Q_3)$ in the EPG model with the experimental data for central Pb-Pb collisions at
$\sqrt{s_{NN}}=2.76$~TeV \cite{ALICE-PRC16}, in the transverse-momentum intervals $0.3<
K_{T3}<1$~GeV/$c$ and $0.16<K_{T3}<0.3$~GeV/$c$.  We first examine the three-pion correlation
functions in the high transverse-momentum interval as shown in Figs.~\ref{zf4-3p-C3exp}(a)
and \ref{zf4-3p-C3exp}(b).  In this momentum interval, the correlation function $C_3(Q_3)$
for the EPG source is almost independent of source particle number $N$.  It is insensitive
to source condensation because most of the pions with high momenta are emitted chaotically
from exited states.  As discussed in the last section, the strength of the multi-pion
correlations for the EPG source varies with temperature in the high momentum interval due to
the variation of the source spatial distribution at different temperatures \cite{LiuRuZhangWong14}.
We determine the temperature $T=$ 100 and 90 MeV for the sources with $C_1=$ 0.35 and 0.40,
respectively, by comparing the calculated three-pion correlation functions with the experimental
data for Pb-Pb collisions at $\sqrt{s_{NN}}=2.76$~TeV at the LHC \cite{ALICE-PRC16}.
Then, we examine the three-pion correlation functions in the low transverse-momentum interval
as shown in Figs.~\ref{zf4-3p-C3exp}(c) and \ref{zf4-3p-C3exp}(d).  In this momentum interval,
the correlation function $C_3(Q_3)$ for the EPG source is sensitive to the source condensation.
Its strength decreases with increasing $N$ because the condensation fraction of source
increases with $N$.  One can see from Fig.~\ref{zf4-3p-C3exp}(d) that the
experimental data are almost between the results for $N=$ 1200 and 1600 for the EPG sources
with $C_1=0.40$, although they are slightly lower than the model results in large $Q_3$
region.  However, the results for the EPG sources with $C_1=0.35$ [Fig.~\ref{zf4-3p-C3exp}(c)]
are higher than the experimental data in large $Q_3$ region.  This may be because the average
longitudinal momentum of the three pions, $K_{L3}=|\textbf{\emph{p}}_{L1}+\textbf{\emph{p}}_{L2} +\textbf{\emph{p}}_{L3}|/3$, in the spherical EPG model is smaller than that in the experiment
in the low transverse-momentum interval.  The values of the three-pion correlation
functions would decrease if we only increase the longitudinal momenta of the pions by a factor
and let all other aspects remain the same.

\begin{figure}[htb]
\vspace*{3mm}
\includegraphics[width=0.49\columnwidth]{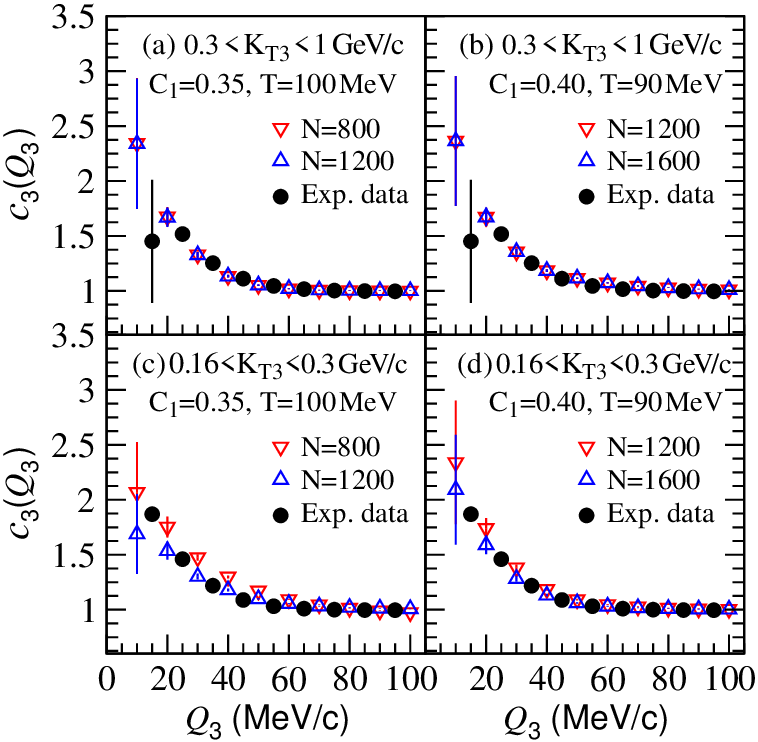}
\vspace{-3mm}
\caption{(Color online) Comparison of the three-pion cumulant correlation function
$c_3(Q_3)$ for the EPG sources and the experimental data for Pb-Pb collisions at
$\sqrt{s_{NN}}=2.76$~TeV \cite{ALICE-PRC16}. }
\label{zf5-3p-c3exp}
\end{figure}

In Fig.~\ref{zf5-3p-c3exp} we show the comparison of the three-pion cumulant correlation
functions $c_3(Q_3)$ in the EPG model with the experimental data for central Pb-Pb
collisions at $\sqrt{s_{NN}}=2.76$~TeV \cite{ALICE-PRC16}, in the transverse-momentum
intervals $0.3<K_{T3}<1$~GeV/$c$ and $0.16<K_{T3}<0.3$~GeV/$c$.  One can see that the
$c_3(Q_3)$ results for the EPG sources are almost consistent with the experimental data
in the high and low transverse-momentum intervals except for those for the EPG source
with $C_1=0.35$ and $N=800$ in the low transverse-momentum interval
[Fig.~\ref{zf5-3p-c3exp}(c)].

\subsection{Four-pion correlations}
\begin{figure}[htb]
\includegraphics[width=0.49\columnwidth]{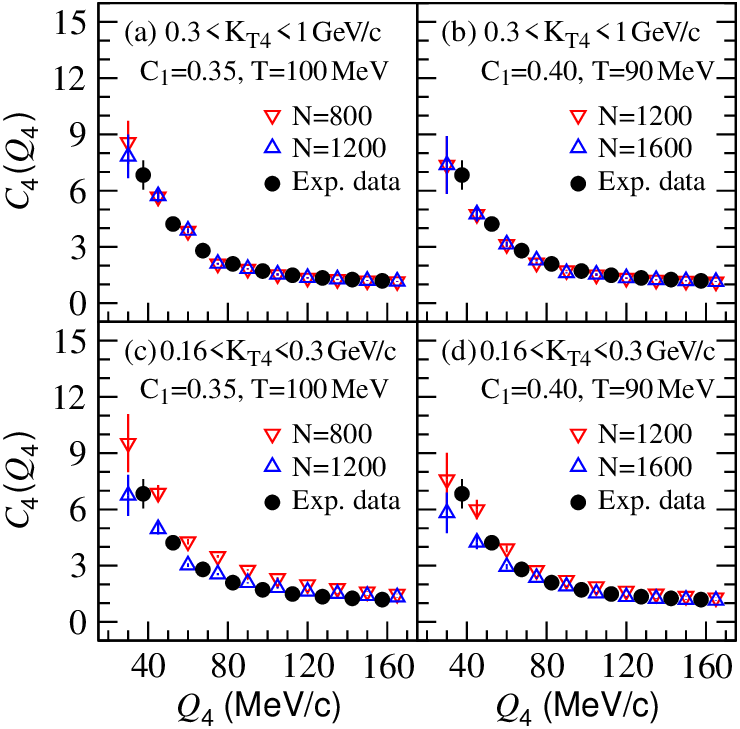}
\vspace*{-3mm}
\caption{(Color online) Comparison of the four-pion correlation function $C_4(Q_4)$
for the EPG sources and the experimental data for Pb-Pb collisions at $\sqrt{s_{NN}}
=2.76$~TeV \cite{ALICE-PRC16}. }
\label{zf4-4p-C4exp}
\end{figure}

We show in Fig.~\ref{zf4-4p-C4exp} the comparisons of the four-pion correlation functions
$C_4(Q_4)$ in the EPG model with the experimental data for central Pb-Pb collisions
at $\sqrt{s_{NN}}=2.76$~TeV \cite{ALICE-PRC16}.  The temperatures of the EPG sources
with $C_1=$ 0.35 and 0.40 are taken to be 100 and 90~MeV, respectively.  One can see from
Figs.~\ref{zf4-4p-C4exp}(a) and \ref{zf4-4p-C4exp}(b) that the the four-pion correlation
functions for the EPG sources are almost independent of the source particle number $N$
because they are insensitive to source condensation in the high momentum interval as
discussed in the last section.  The model results are consistent with the experimental
data in the high transverse-momentum interval as the three-pion correlation functions.
From Figs.~\ref{zf4-4p-C4exp}(c) and \ref{zf4-4p-C4exp}(d) one can see that the four-pion
correlation function for the EPG source with larger $N$ is lower than that for the source
with smaller $N$ in the low transverse-momentum interval.  It is because the condensation
fraction is higher for the source with larger $N$.  In small $Q_4$ region, the experimental
data are between the model results for the sources with the small and large $N$.  In large
$Q_4$ region, the model results are slight higher than the experimental data.  This may be
because the average longitudinal momentum of the four pions, $K_{L4}=|\textbf{\emph{p}}_{L1}
+\textbf{\emph{p}}_{L2} +\textbf{\emph{p}}_{L3}+\textbf{\emph{p}}_{L4}|/4$, in the spherical
EPG model is smaller than that in experiment in the low transverse-momentum interval.

\begin{figure}[htb]
\includegraphics[width=0.48\columnwidth]{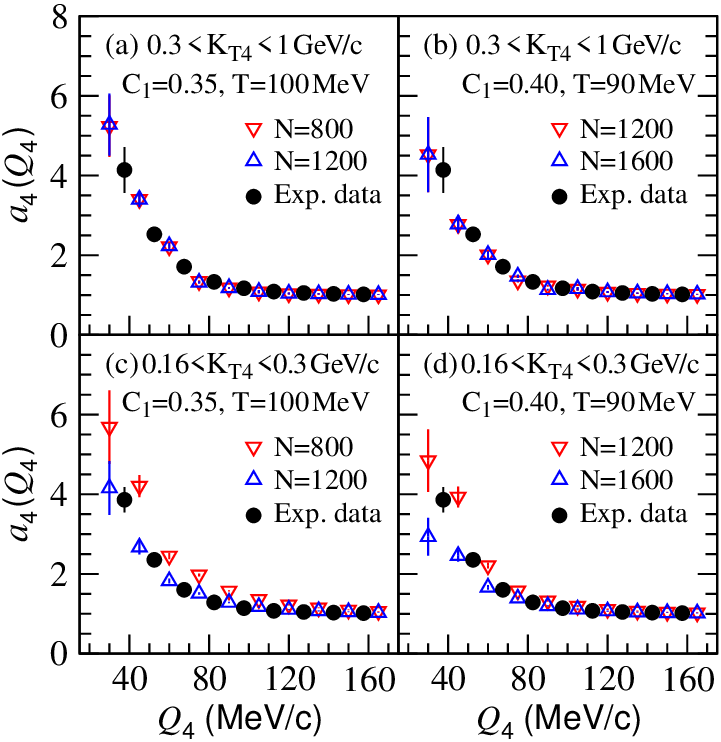}
\vspace*{-3mm}
\caption{(Color online) Comparison of the four-pion cumulant correlation function
$a_4(Q_4)$ for the EPG sources and the experimental data for Pb-Pb collisions at
$\sqrt{s_{NN}}=2.76$~TeV \cite{ALICE-PRC16}. }
\label{zf5-4p-a4exp}
\end{figure}

\begin{figure}[htb]
\includegraphics[width=0.48\columnwidth]{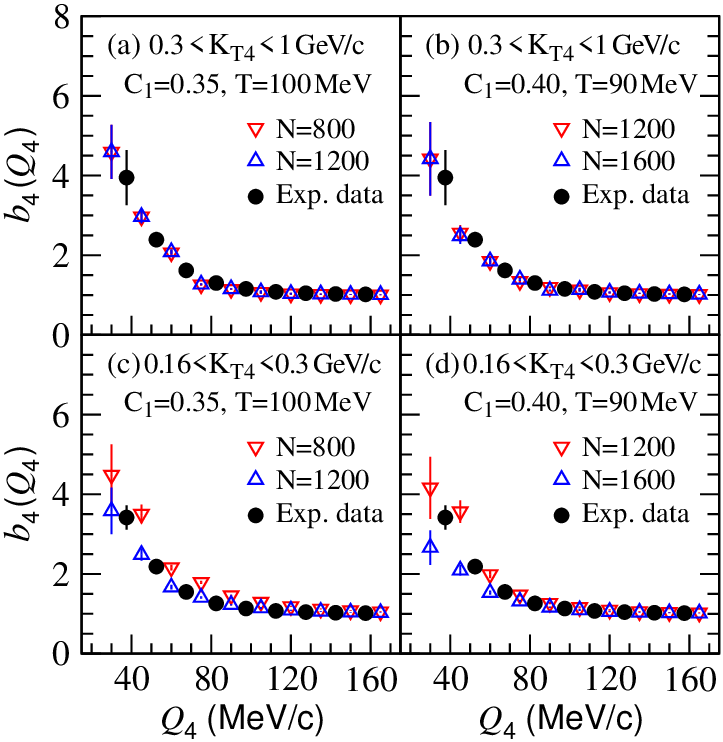}
\vspace*{-3mm}
\caption{(Color online) Comparison of the four-pion cumulant correlation function
$b_4(Q_4)$ for the EPG sources and the experimental data for Pb-Pb collisions at
$\sqrt{s_{NN}}=2.76$~TeV \cite{ALICE-PRC16}. }
\label{zf5-4p-b4exp}
\end{figure}

We show in Figs.~\ref{zf5-4p-a4exp} and \ref{zf5-4p-b4exp} the comparisons of the
four-pion cumulant correlation functions $a_4(Q_4)$ and $b_4(Q_4)$ in the EPG model
with the experimental data for central Pb-Pb collisions at $\sqrt{s_{NN}}=2.76$~TeV
\cite{ALICE-PRC16}, respectively.  Because the correlations of single pair are
removed from $a_4$ and the correlations of single and double pair are removed from
$b_4$, the results of $a_4(Q_4)$ are lower than those of $C_4(Q_4)$, and the results
of $b_4(Q_4)$ are further lower than those of $a_4(Q_4)$.  One can see that the model
results of $a_4(Q_4)$ and $b_4(Q_4)$ in the high transverse-momentum interval are
almost independent of the source particle number $N$.  They are consistent with the
experimental data in the high momentum interval.
However, the model results are sensitive to the source particle number $N$ in the low
transverse-momentum interval.  The experimental data are almost between the model
results for $N=$ 1200 and 1600 for the sources with $C_1=0.40$ and more consistent
with the model results for $N=1200$ for the source with $C_1=0.35$.

\begin{figure}[htb]
\vspace*{3mm}
\includegraphics[width=0.49\columnwidth]{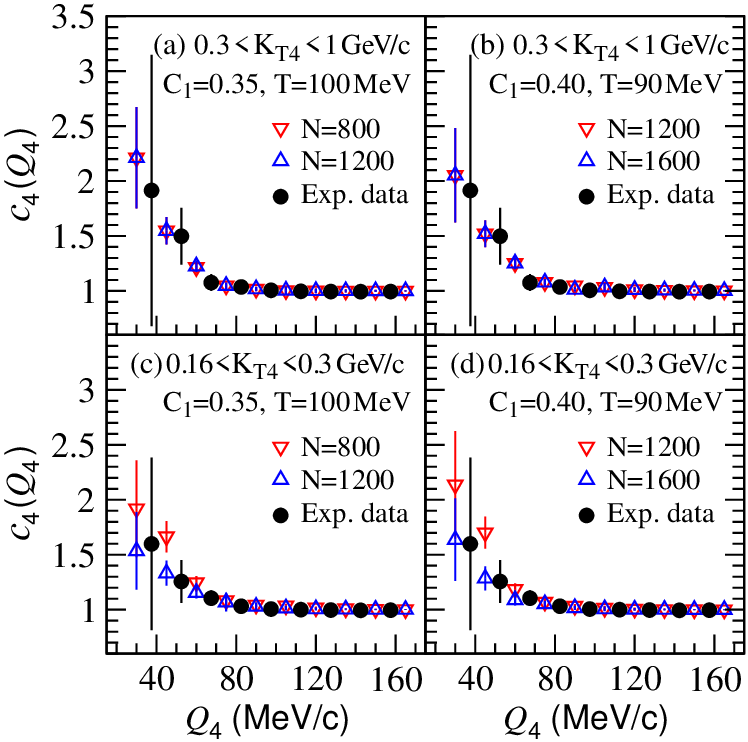}
\vspace{-3mm}
\caption{(Color online) Comparison of the four-pion cumulant correlation function
$c_4(Q_4)$ for the EPG sources and the experimental data for Pb-Pb collisions at
$\sqrt{s_{NN}}=2.76$~TeV \cite{ALICE-PRC16}. }
\label{zf5-4p-c4exp}
\end{figure}

We show in Fig.~\ref{zf5-4p-c4exp} the comparisons of the four-pion cumulant correlation
functions $c_4(Q_4)$ in the EPG model with the experimental data for central Pb-Pb
collisions at $\sqrt{s_{NN}}=2.76$~TeV \cite{ALICE-PRC16}.  It should be mentioned that
the error bars of the experimental data shown in the figures in this paper are statistic
error plus the system error, which is large for $c_4(Q_4)$ in the small $Q_4$ region
\cite{ALICE-PRC16}.  The error bars of the model results shown in the figures in this
paper are statistic error.  One can see that the model results of $c_4(Q_4)$ are
independent of the particle number of the sources and consistent with the experimental
data in the high transverse-momentum interval.
However, the model results of $c_4(Q_4)$ in the low transverse-momentum interval are
particle-number dependent in the low $Q_4$ region.  The experimental data of $c_4(Q_4)$
in the low transverse-momentum interval are between the model results for the low and
high $N$ in the small $Q_4$ region, and can be reproduced by the EPG model in the large
$Q_4$ region.

\subsection{Condensation fraction}
\begin{figure}[htb]
\vspace*{5mm}
\includegraphics[width=0.80\columnwidth]{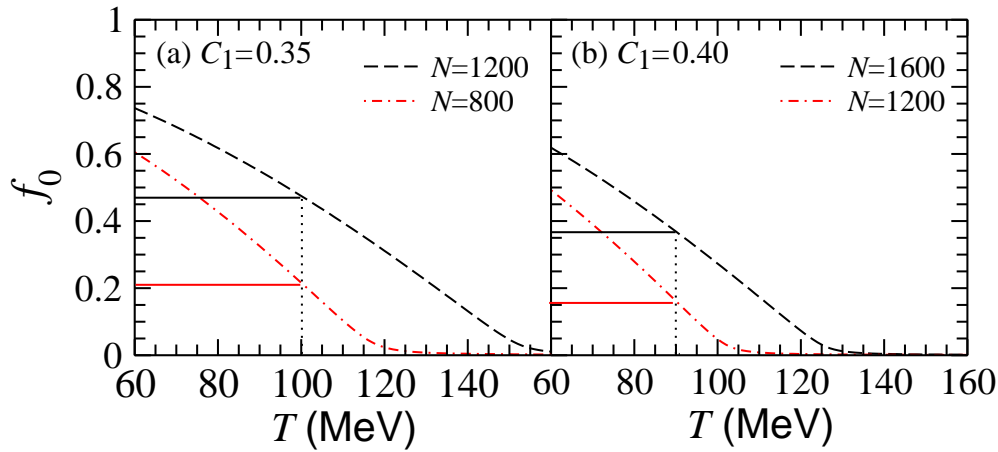}
\vspace{-3mm}
\caption{(Color online) Condensation fractions for the EPG sources with $T=100$~MeV for
$C_1=0.35$ and $T=90$~MeV for $C_1=0.40$. }
\label{zf6-f0}
\end{figure}

We find in the last two subsections that the three- and four-pion correlation functions
in the EPG model can reproduce in some degree the experimental data for Pb-Pb collisions
at $\sqrt{s_{NN}}=2.76$~TeV \cite{ALICE-PRC16}.
By comparing with the experimental data, we determine that the most suitable temperatures
for the EPG sources with $C_1=$ 0.35 and 0.40 are 100 and 90~MeV, and the particle numbers
are perhaps in the regions $[800,1200]$ for the source with $C_1=0.35$ and $[1200,1600]$
for the source with $C_1=0.40$.  With these source parameters, we further determine the
condensation fractions between 0.22~--~0.47\% for the source with $C_1=0.35$ and
0.16~--~0.37\% for the source with $C_1=0.40$, as shown in Fig. \ref{zf6-f0}.

In Ref. \cite{ALICE-PRC16}, the ALICE collaboration extracted the coherent faction,
32\% $\pm$ 3\%(stat) $\pm$ 9\%(syst), by analyzing the suppression of four-pion
correlations in Pb-Pb collisions at $\sqrt{s_{NN}}=2.76$~TeV, and pointed out
that this coherent fraction value cannot explain the suppression of three-pion
correlations observed.  In the EPG model, the pion emission from ground state is
coherent and the condensation fraction defined by Eq. (\ref{def-f0}) is coherent
fraction.  The values of condensation fraction are determined by the comparisons
of the model results and experimental data of three- and four-pion correlations.
They are consistent with the value of coherent fraction extracted by the ALICE
collaboration \cite{ALICE-PRC16}.
Also, in the EPG model the source size for a smaller $C_1$ parameter is smaller
than that for a larger $C_1$ parameter.  Considering the source size is larger
for central collisions than that for peripheral collisions in experiments, the
determined condensation fraction 0.22~--~0.47\% for $C_1=0.35$ and 0.16~--~0.37\%
for $C_1=0.40$ are also consistent with the conclusion of experimental analyses
that ``There does not appear to be a significant centrality dependence to the
extracted coherent fractions." \cite{ALICE-PRC16}
According to the EPG model, the condensation not only depends on the particle
number which is smaller in peripheral collisions than in central collisions, but
also depends on the source size which is also smaller in peripheral collisions
than in central collisions.  The condensation degree increases with increasing
particle number and decreases with increasing source size.  So, the comprehensive
effect of particle number and source size may lead to the result that the condensation
fraction or coherent fraction is independent of collision centrality.

Finally, it should be mentioned that the EPG model deals with the canonical ensemble
in which the systems of pion gas have a fixed particle number $N$.  So, the two- and
multi-pion correlation functions calculated in the EPG model are the so-called
``exclusive correlation functions" \cite{CsorgoZimanyi97}.  They should be compared
with the corresponding experimental correlation functions obtained from the events
with the same multiplicity.  However, because of data statistics the correlation
functions obtained experimentally are from many events in some multiplicity intervals.
In this case, a strict comparison should be between the experimental data in a
multiplicity interval and the averaged EPG exclusive results over the same multiplicity
interval with the weights of multiplicity obtained experimentally.  On the other hand,
it is also meaningful to make a comparison between the experimental correlation functions
in a multiplicity interval and the EPG exclusive results with the particle number
consistent with the average multiplicity in the multiplicity interval, if the differences
between the exclusive and inclusive correlation functions are negligible approximately.
In fact, the the difference between the inclusive and exclusive correlation functions
is from the effects of higher-order correlations \cite{CsorgoZimanyi97}, the residual
correlation effects in single- two- and multi-pion samples \cite{WAZajc84,Chacon91,Zhang9390}.
In a $m$-pion sample, the leading-order effect of multi-pion correlations is approximately
proportional to $\big[m\!\cdot\!\int |{\tilde \rho}(\textbf{\emph{p}},E(\textbf{\emph{p}}))|
d^3p\big]$, where $\tilde \rho(p)$ is the on-shell Fourier transform of source density,
which is very small in high-energy heavy-ion collisions where the source radius and lifetime
are about 10 fm and 10 fm/$c$ \cite{Gyu79,CsorgoZimanyi97}.  More detailed investigations of
the difference between the exclusive and inclusive correlation functions and the comparison
between the EPG correlation functions and the experimental data will be of great interest.
Additionally, the intercepts of pion HBT correlation functions can be affected by long-lived 
resonance decays, their effects on pion transverse-momentum spectra are discussed in the 
chemical nonequilibrium thermal model \cite{Begun14,Begun15}.  It will be of considerable 
interest to estimate the influence of long-lived resonance decay and remove the influence 
in the coherence analyses of multi-pion interferometry.

\section{Summary and conclusion}
We have calculated the three- and four-pion correlations in the EPG model with
Bose-Einstein condensation.  The relationship between the multi-pion correlations
and the source condensation fraction is investigated.  It is found that the multi-pion
correlation functions and cumulant correlation functions are sensitive to the
condensation fraction of the EPG source in the low transverse-momentum intervals
of the three and four pions, $K_{T3,T4}<0.3$~GeV/$c$.  These correlation functions
exhibit significant decreases with decreasing source temperature and increasing source
particle number in the low transverse-momentum intervals, because the condensation
fraction of the EPG source is high at a low temperature and large particle number.
On the other hand, the multi-pion correlation functions and cumulant correlation
functions are insensitive to the source condensation in the high transverse-momentum
intervals $K_{T3,T4}>0.3$~GeV/$c$.  They are almost independent of the source particle
number in the high transverse-momentum intervals, because most of the pions with high
momenta are emitted chaotically from excited states in the EPG model even if with a
considerable condensation fraction.  We have compared the model results of three- and
four-pion correlation functions and cumulant correlation functions with the experimental
data for Pb-Pb collisions at $\sqrt{s_{NN}}=2.76$~TeV at the LHC.  It is found that
the multi-pion correlation functions and cumulant correlation functions in the EPG
model may reproduce the experimental results in a considerable degree.  The source
condensation fraction determined by the comparisons is between 16~--~47\%.
Further investigations of the comparison between the EPG correlation functions and the
experimental data are of great interest.

\begin{acknowledgments}
We thank Cheuk-Yin Wong for helpful discussions.
This research was supported by the National Natural Science Foundation of China under
Grant Nos. 11675034 and 11275037, and the China Scholarship Council.
\end{acknowledgments}

\end{document}